\documentclass{aa}  
\usepackage{graphicx}
\usepackage{txfonts}
\usepackage[colorlinks=true, linkcolor=blue, urlcolor=blue, citecolor=blue]{hyperref}
\usepackage{color}

\begin{document}

   \title{The multi-thermal chromosphere:}

   \subtitle{inversions of ALMA and IRIS data}

   \author{J. M. da Silva Santos\inst{1} \and J. de la Cruz Rodríguez\inst{1} \and J. Leenaarts\inst{1} \and G. Chintzoglou\inst{2,3} \and B. De Pontieu\inst{2,4,5} \and\\ S. Wedemeyer\inst{4,5} \and M. Szydlarski\inst{4,5}
          }

   \institute{Institute for Solar Physics, Department of Astronomy, Stockholm University, AlbaNova University Centre, SE-106 91 Stockholm, Sweden, \email{joao.dasilva@astro.su.se}
         \and
            Lockheed Martin Solar \& Astrophysics Laboratory, 3251 Hanover St, Palo Alto, CA 94304, USA
         \and
            University Corporation for Atmospheric Research, Boulder, CO 80307-3000, USA
         \and
            Rosseland Centre for Solar Physics, University of Oslo, P.O. Box 1029 Blindern, NO0315 Oslo, Norway
        \and
            Institute of Theoretical Astrophysics, University of Oslo, P.O. Box 1029 Blindern, NO0315 Oslo, Norway\\
             }

   \date{}

% \abstract{}{}{}{}{} 
% 5 {} token are mandatory
 
  \abstract
   {Numerical simulations of the solar chromosphere predict a diverse thermal structure with both hot and cool regions. Observations of plage regions in particular typically feature broader and brighter chromospheric lines, which suggests that they are formed in hotter and denser conditions than in the quiet Sun, but also implies a nonthermal component whose source is unclear.} 
   {We revisit the problem of the stratification of temperature and microturbulence in plage and the quiet Sun, now adding millimeter (mm) continuum observations provided by the Atacama Large Millimiter Array (ALMA) to inversions of near-ultraviolet Interface Region Imaging Spectrograph (IRIS) spectra as a powerful new diagnostic to disentangle the two parameters. We fit cool chromospheric holes and track the fast evolution of compact mm brightenings in the plage region.}
   {We use the STiC nonlocal thermodynamic equilibrium (NLTE) inversion code to simultaneously fit real ultraviolet and mm spectra in order to infer the thermodynamic parameters of the plasma.}
   {We confirm the anticipated constraining potential of ALMA in NLTE inversions of the solar chromosphere. We find significant differences between the inversion results of IRIS data alone compared to the results of a combination with the mm data: the IRIS+ALMA inversions have increased contrast and temperature range, and tend to favor lower values of microturbulence ($\sim$3-6$\rm\,km\,s^{-1}$ in plage compared to $\sim$4-7$\rm\,km\,s^{-1}$ from IRIS alone) in the chromosphere. The average brightness temperature of the plage region at 1.25\,mm is 8500\,K, but the ALMA maps also show much cooler ($\sim3000$\,K) and hotter ($\sim11\,000$\,K) evolving features partially seen in other diagnostics. To explain the former, the inversions require the existence of localized low-temperature regions in the chromosphere where molecules such as CO could form. The hot features could sustain such high temperatures due to non-equilibrium hydrogen ionization effects in a shocked chromosphere -- a scenario that is supported by low-frequency shock wave patterns found in the \ion{Mg}{II} lines probed by IRIS.}
   {}


   \maketitle
%
% ------------------------------------------------------------------------------ %
\section{Introduction}
\label{section:introduction}

Plage are bright patches in the solar atmosphere that are associated with concentrations of predominantly unipolar, vertical magnetic fields. Their study is important because they offer a way of investigating the interplay between magnetic flux emergence, dynamics, and energy deposition and transport to higher layers \citep[e.g.,][]{1974SoPh...39...49S,1992ApJ...393..782T,1997ApJ...474..810M,2012A&A...545A..22K,2016A&A...589A...6K,2015A&A...576A..27B,2019arXiv190807464B,2003ApJ...590..502D,2015ApJ...799L..12D,2017ApJS..229....4C}. 

\citet{Carlsson_2015} reported on observations of a plage region in the ultraviolet (UV) with the Interface Region Imaging Spectrograph \citep[IRIS,][]{2014SoPh..289.2733D}, and showed that the chromospheric \ion{Mg}{ii} h and k resonance lines are brighter and broader, and their central reversals are shallower (or absent) compared to quiet-Sun (QS) profiles, whereas the subordinate UV triplet lines \citep[diagnostics of the low chromosphere, ][]{Pereira15} are usually in absorption, suggesting that plage have a hot ($\sim$\,6000-6500\,K) and dense chromosphere. 
Furthermore, the temperature of the chromospheric plateau and its location in height, in addition to microturbulence, were shown to have an effect on the core width and intensity of plage profiles. 
 
This picture is in agreement with non-local thermodynamic equilibrium (NLTE) inversions of \ion{Mg}{II} plage profiles, which infer enhanced chromospheric temperatures of the same order, and microturbulence velocities up to $\sim$\,$8\,\rm km\,s^{-1}$ in the line-forming region \citep{2016ApJ...830L..30D}, which is consistent with the nonthermal widths of the optically thin line of \ion{O}{I} \citep{Carlsson_2015}. An additional broadening of the order of $\sim$1-3\,$\rm km\,s^{-1}$ has also been shown to be necessary to fit photospheric \ion{Fe}{I} lines observed towards plage compared to the QS \citep{2019arXiv190807464B}.
The source of the high microturbulence remains unclear, but possible explanations may reside, for example, in small-scale Alfvén waves and magneto-acoustic shocks from below \citep{2015ApJ...799L..12D} or Alfvénic turbulence \citep{2011ApJ...736....3V}. We refer to \citet{2019ARA&A..57..189C} for a more complete review of heating mechanisms in plage regions.

Plage regions are also known to be brighter than the QS in the millimeter (mm) continuum \citep[e.g.,][]{1962SvA.....6..202S,1970SoPh...13..348K,1971SoPh...21..130K,1971SoPh...16...87B,1995ApJ...453..511L}. Observations in the mm provide valuable insight into the temperature stratification of plage because the source function of the free-free continuum has an almost linear dependence on the gas temperature in the chromosphere \citep[see review by][]{Wedemeyer16}, and is insensitive to the microturbulence parameter. However, the opacities may be controlled by time-dependent hydrogen ionization associated with shocks \citep{2002ApJ...572..626C,2006ASPC..354..306L,2007A&A...471..977W}, which are frequent in plage regions \citep[e.g.,][and references therein]{2007ASPC..368...65D,2019ARA&A..57..189C}.

The Atacama Large Millimeter/Submillimeter Array \citep[ALMA,][]{2009IEEEP..97.1463W} has overcome the spatial resolution limitations that observing at such long wavelengths entails. ALMA has shown that the landscape of the solar surface at 1\,mm is marked by relatively cooler regions such as the QS and sunspot umbras, and hotter features such as plage regions \citep{2017ApJ...850...35L,Brajsa2018}, and that the brightness of the mm continuum correlates with that of the \ion{Mg}{II} lines \citep{2017ApJ...845L..19B,2018ApJ...860L..16B,2019A&A...622A.150J}. 
However, the large scatter and offset between the two diagnostics is evidence for NLTE effects in the \ion{Mg}{II} lines that make them only partially sensitive to the chromospheric temperatures \citep[e.g.,][]{2013ApJ...772...90L}, although there may also be systematic differences in their formation heights \citep[e.g.,][]{PaperI}.
This problem may play a role in the discovery of "chromospheric holes" in 3\,mm ALMA maps. These features have brightness temperatures a few thousand kelvin lower than the QS average and appear to be invisible in \ion{Mg}{II}\,k and \ion{H}{$\alpha$} broadband images \citep{2019ApJ...877L..26L}.  

Based on a snapshot of a 3D radiation-magneto-hydrodynamics (r-MHD) simulation of the solar atmosphere, \citet{PaperI} showed that mm observations help to constrain temperatures in inversions of synthetic NLTE lines such as the \ion{Mg}{II} h and k doublet. Here we report on the results of the first inversions of real active region plage data obtained during a coordinated IRIS and ALMA campaign. We present constraints on temperature and microturbulence of plage, and we demonstrate the synergy between IRIS and ALMA, using this to infer the atmospheric parameters of both cool and hot chromospheric regions.

\begin{figure*}
    \centering
    \includegraphics[width=\linewidth]{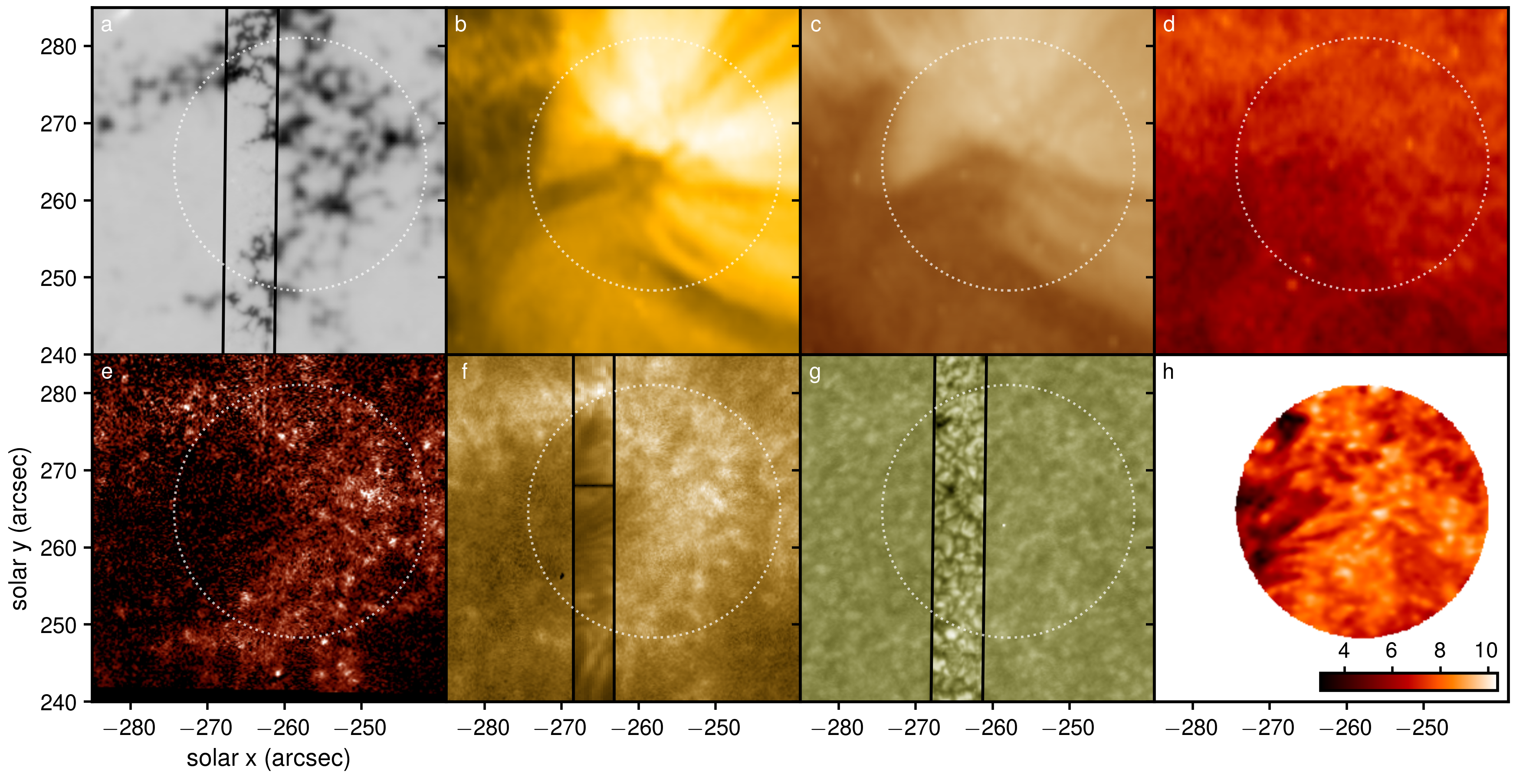}
    \caption{{ }Overview of the target on the 22nd April 2017 as observed by SDO, IRIS, Hinode, and ALMA. {\it Panel (a):} SDO/HMI magnetogram on the background (16:00\,UT) and Hinode/SOT/SP magnetogram (15:58:11-15:59:35\,UT) overplotted; {\it Panel (b):} SDO/AIA\,171\,\AA~(15:59:09\,UT); {\it Panel (c):} AIA\,193\,\AA~(15:59:16\,UT); {\it Panel (d):} AIA\,304\,\AA~(15:59:17\,UT); {\it Panel (e):} IRIS SJI\,1440\,\AA~(15:59:15\,UT); {\it Panel (f):} IRIS SJI\,2796\,\AA~on the background (15:59:09\,UT) and raster scan in the k core overplotted; {\it Panel (g):} IRIS SJI\,2832\,\AA~(15:59:18\,UT) on the background and Hinode/SOT/SP 6302\,\AA~continuum (15:59:35\,UT) overplotted; {\it Panel (h):} ALMA 1.25\,mm (15:59:11\,UT) brightness temperature in units of kK. The dotted circles correspond to the ALMA FoV. The AIA images are displayed on a logarithmic scale.} 
    \label{fig:overview}
\end{figure*}

% ------------------------------------------------------------------------------ %
\section{Observations}

\subsection{Data reduction and alignment}
\label{section:observations}

The observations were carried on 22 April 2017, when ALMA and IRIS co-observed part of a plage on the eastern side of the active region NOAA\,12651 at $\mu\approx0.92$ (where $\mu$ is cosine of the heliocentric angle) between 15:59 UT and 16:34 UT. For context, we also use images taken with the Atmospheric Imaging Assembly \citep[AIA,][]{2012SoPh..275...17L} and the line-of-sight magnetograms taken with the Helioseismic and Magnetic Imager \citep[HMI,][]{2012SoPh..275..207S} on board the Solar Dynamics Observatory (SDO), and Hinode/SOT/SP \citep{2008SoPh..249..167T} slit scans of the \ion{Fe}{I}\,6301 and 6302\,\AA~lines.  

The IRIS data consist of dense (0.349\arcsec) 16-step rasters with a pixel size of 0.167\arcsec\, along the slit, scanning a field of view (FoV) of $5.3\arcsec\times66.7\arcsec$ initially centered at the edge of the plage at $[x,y]\sim$ $[-265\arcsec, 269\arcsec]$ where the $x$ and $y$ axes are the helioprojective longitude and latitude, respectively. The exposure time per step was 0.5\,s and the raster cadence was 25\,s. The data have been flat-fielded, dark-subtracted, corrected for geometrical distortions, and calibrated in wavelength as described in \citet{2018SoPh..293..149W}.
The radiometric calibration was performed using version four of the calibration files obtained using the \textit{iris\_get\_response} routine in SolarSoft \citep[SSW, ][]{1998SoPh..182..497F} with a flux calibration uncertainty of 15\%. The noise level (rms) is approximately $\delta I\sim1\times10^{-7}\,\rm erg\,s^{-1} cm^{-2} sr^{-1} Hz^{-1}$. We also used IRIS slit-jaw-images (SJI) for context and alignment purposes. 

The Hinode/SOT rasters were downloaded from the Community Spectro-polarimetric Analysis Center (CSAC) page\footnote{\url{http://www2.hao.ucar.edu/csac}} as level 2 data, which are the results of Milne-Eddington inversions with the MERLIN code \citep[see calibration details in][]{2013SoPh..283..601L}. The step length is 0.148\arcsec~and the pixel scale is 0.159\arcsec~along the slit. The exposure time was 1.6\,s. The alignment of the Hinode and IRIS data was verified by matching the 6302\,\AA~continuum with an average of several IRIS SJI\,2832\,\AA~images within the Hinode raster time span. Unfortunately, Hinode stopped observing a few seconds after the start of the ALMA campaign, so we can only use the last raster that is closest in time (15:58:11-15:59:35\,UT) for context.

The calibrated ALMA Band 6 data (as downloaded from the ALMA Archive\footnote{\url{http://almascience.nrao.edu/asax/}}) consist of five interferometric scans \citep{2017SoPh..292...87S} and five single-dish scans of the full solar disk with Total Power (TP) antennas \citep{2017SoPh..292...88W}. The net interferometric observing time on target spans 37\,min divided into five scans of which the first four have durations of about 8\,min each, separated by roughly 2 min calibration breaks; these are carried out within the above-mentioned ALMA-IRIS co-observing window. 
The data are processed with the Solar ALMA Pipeline (SoAP2, Szydlarski et al. in prep), which includes self-calibration, deconvolution of the synthesized beam of ALMA (i.e., the PSF of the interferometric array) via application of the multi-scale CLEAN algorithm by \citet{2011A&A...532A..71R} as implemented in CASA \citep{2007ASPC..376..127M}, primary beam correction, and combination of the interferometric data with the TP data. The latter provide the absolute scale for the derived brightness temperatures. Following the recommendation by \citet{2017SoPh..292...88W}, the absolute scale is corrected by rescaling the TP maps based on the average value in the disk center region to a reference value of 5900\,K before combining with the interferometric data. Multi-frequency synthesis is employed in order to increase image fidelity, resulting in one brightness temperature map at 2\,s cadence corresponding to the central frequency of Band 6 and thus to an effective wavelength of 1.25 mm. We refer to \citet{Wsubmitted} for more details about data processing with SoAP.

The largest source of uncertainty remains the calibration of the TP maps, which results in an uncertainty in the absolute temperature scale. According to the Proposer’s Guide\footnote{\url{https://almascience.nrao.edu/documents-and-tools/cycle4/alma-proposers-guide}} for Cycle 4, the absolute calibration of the single-dish TP brightness temperatures has an uncertainty of bettween 10 and 15\%, which translates to 590\, to 885\,K based on the reference value of $5900$\,K recommended by \citet{2017SoPh..292...88W}.  However, the analysis of ALMA data implies that the uncertainty might be as low as 5\% or even less \citep[e.g.,][]{Rodger_2019,2019A&A...622A.150J}. In addition to the uncertainty in the TP calibration, additional uncertainties in the interferometric brightness temperature differences have to be considered but they are expected to be smaller. 
The noise level is $\sim30$\,K at most. While a systematic error analysis for the derived brightness temperatures has yet to be carried out, we suggest that the combined uncertainty of the absolute brightness temperatures is likely on the order of 3-5\% but we cannot rule out potentially larger uncertainties.

The ALMA and IRIS data were aligned and interpolated to a common grid. Up to 14 ALMA images were then combined to replicate the IRIS rasters, meaning that the time lag between each IRIS slit and ALMA map is 2\,s at most. Finally, the ALMA brightness temperatures were converted to intensities via the Planck's function as required for the inversions described in Section\,\ref{section:setup}.

\subsection{Overview}
\label{section:overview}
Figure\,\ref{fig:overview} shows an overview of the region of interest as seen by SDO, Hinode, IRIS, and ALMA (one time stamp). 
Throughout the ALMA time series we find brightness temperatures reaching up to $\sim12\,500$\,K in relatively compact ($\sim1-2$\arcsec) brightenings with temperature variations of 1000 K over timescales of a few minutes, some of which coincide with the IRIS \ion{Si}{IV} and \ion{Mg}{II} brightenings, but also low temperatures down to $\sim3000$\,K -- cooler than the classical temperature minimum -- with no clear counterparts in the IRIS SJIs. The average value for the center of the plage region is $\sim8500$\,K, which is far higher than the typical $\sim6000-6500$\,K brightness temperature of the peaks of the \ion{Mg}{II} h and k lines. For comparison, the average quiet-Sun value at disk center at 1.3\,mm is $\sim5900$\,($\pm100$) K \citep{2017SoPh..292...88W}. The ALMA maps also show long, hot loops emanating from the plage to the quieter areas, analogous to the bright loops observed at 3\,mm \citep{2019arXiv190608896M}.

The coolest region within the FoV of the IRIS raster has an average $T_{\rm b}\sim4050$\,K and a minimum $T_{\rm b}\sim3100$\,K. It is cospatial with a small ($\sim$2\arcsec$\times$2\arcsec) pore (yet slightly more extended) at the edge of the ALMA FoV that is clearly seen in the 6302\,\AA~continuum and, to a lesser extent, in IRIS SJI\,2832\,\AA~(Fig.\,\ref{fig:overview}). \citet{2019ApJ...877L..26L} recently found a "chromospheric hole" with a size of 20\arcsec$\times$20\arcsec~in ALMA QS data taken in Band\,3 at 3\,mm, but it is unclear whether they are similar phenomena because of their different size and location. Whereas the 3\,mm holes in the QS seem to be unrelated to the photospheric magnetic field, the 1.25\,mm hole presented here is coincidentally hovering above a $\sim$1.3-1.8\,kG magnetic concentration with a field inclination of $\sim$160-170$^{\circ}$~as measured by Hinode. The range in the HMI magnetogram is approximately [-840, 230]\,G. Interestingly, its brightness temperatures are even lower than the $\sim5300$\,K emission above a sunspot umbra with a magnetic field reaching over $\sim$2.5\,kG \citep{2017ApJ...850...35L}. 

As in \citet{2019ApJ...877L..26L} the chromospheric hole has no obvious signature in the 2796\,\AA~SJI images, but the raster scans clearly show a dark region at that location, albeit only slightly darker than the canopy (Fig.\,\ref{fig:overview}). 
The other cool areas in the ALMA maps do not have an obvious photospheric counterpart judging by the IRIS SJIs and HMI continuum. They may trace the cool parts of the loops partly seen in the \ion{Mg}{II} lines, as far as we can tell from the limited coverage of the IRIS raster and the low contrast of the 2796\,\AA~SJIs.

We do not find any single peaked \ion{Mg}{II} profiles in the plage region scanned by the IRIS slit but they indeed show shallower central reversals compared to the weakly magnetized areas as reported by \citet{Carlsson_2015}.
The AIA coronal channels do not show any coronal analog of the chromospheric holes, and do not reveal flaring activity in the region. 

% ------------------------------------------------------------------------------ %
\section{Inversions of IRIS and ALMA data}
\label{section:results}

Performing NLTE inversions of the 35\,min time series in the whole FoV is very computationally demanding, typically requiring several hours on thousands of parallel cores per inversion cycle per raster, and manual inspection of the goodness of the fit in between. Therefore, we focus only on selected parts of the FoV and limited time span. We report on the inversion results of one full ($\sim$5\arcsec$\times$30\arcsec)~IRIS+ALMA raster taken
at approximately 15:59\,UT, and on the time evolution of a smaller subfield ($\sim$3.5\arcsec$\times$7.5\arcsec)~between 15:59 and 16:07\,UT which features a few interesting transient mm brightenings.

\subsection{Inversion setup}
\label{section:setup}
The inversions were performed with the STockholm Inversion Code\footnote{\url{https://github.com/jaimedelacruz/stic}} \citep[STiC, ][]{2016ApJ...830L..30D,2019A&A...623A..74D,2001ApJ...557..389U} which is a regularized Levenberg-Marquardt code for multi-atom NLTE inversions of multiple spectral lines and continua including partial frequency redistribution (PRD) effects, which are necessary to explain the h and k lines of \ion{Mg}{II} \citep[e.g.,][]{1974ApJ...192..769M,1997SoPh..172..109U,2013ApJ...772...89L}. 

Hydrogen and magnesium ionization are treated in NLTE by solving the statistical equilibrium equation together with charge conservation. This is probably a good approximation for \ion{Mg}{II} \citep{2013ApJ...772...89L}, but it may not be so for hydrogen, whose ionization degree is time-dependent owing to the large ionization and/or recombination timescales compared to the dynamic timescales in the chromosphere \citep[e.g.,][]{2002ApJ...572..626C,2006A&A...460..301L,2006ASPC..354..306L,2017A&A...598A..89R}. Consequently, (time-dependent) non-equilibrium electron and proton densities may affect the opacity of the mm continuum. 
It is not possible to take these effects into account with STiC or with any other current inversion code to our knowledge. Consequently, during the passage of a shock the modeled mm intensities may reflect stronger variations of electron densities, and give larger temperature increments with height than in reality \citep{2002ApJ...572..626C}. We compared inversions with LTE and NLTE hydrogen ionization and verified that the NLTE calculations give us much more uniform electron densities in the chromosphere which are expected to be more realistic according to simulations \citep[e.g.,][]{2006A&A...460..301L,2007A&A...473..625L}. Therefore, we used (instantaneous) NLTE electron densities.

We assume that the sources of opacity in the mm range are free-free processes and neglect cyclotron emission, which would only be significant above magnetic field elements with strengths of several kilo-Gauss at longer mm wavelengths ($\nu_{\rm B}/\nu \lessapprox 2\times10^{-5}$, where $\nu_{\rm B}$ is the Larmor frequency), and synchrotron radiation (or nonthermal gyroemission), which is only expected at flaring sites \citep[e.g.,][]{Wedemeyer16}. None of these mechanisms are relevant in the plage region probed by ALMA judging by the activity level seen in the AIA context images (see Fig.\,\ref{fig:overview}). 

The calibrated data $I_{\rm obs}$ are passed to STiC that iteratively adjusts the plasma parameters\footnote{The magnetic field vector is unconstrained from these data.} {\boldmath $p:$}   in this case, only temperature, line-of-sight velocity, microturbulent velocity, and gas pressure at the boundary ($[T, v_{\rm LOS}, v_{\rm turb}, p^0_{\rm gas}$]) as a function of logarithm of optical depth of the 500\,nm radiation, hereafter $\log\tau$. The synthetic spectra $I_{\rm syn}$ minimizes the $\chi^2$ function defined as follows:
\begin{equation}
    \chi^2({\bf p}) = \frac{1}{n}\sum^{N}_{i=1} \left(\frac{I_{\rm obs}(\nu_{i})-I_{\rm syn}(\nu_{i}, {\bf p})}{\sigma_{i}}\right)^2 + \sum^{N_{\rm p}}_{j=1}\alpha_{j}\,r_{j}({\bf p}) \label{eq:chisq}
,\end{equation}
\noindent where the first term is the mean squared error weighted by $\sigma$ data weights (e.g., uncertainty estimate), and the second (optional) term is a sum of regularization terms which are essentially penalty functions with $\alpha$ being the weights of the regularization for each parameter. For example, the microturbulence is penalized such that the optimization should prefer $v_{\rm turb}(\log\tau)\sim0$ unless it is strictly necessary, which is usually the case in plage regions.

Given the very low statistical weight of the single mm point, we introduced an \textit{ad hoc} weighting function that ensures that the ALMA data are well fitted (within the uncertainties), otherwise its impact on $\chi^2$ (Eq.\,\ref{eq:chisq}) would be negligible. 
This comes with the potential risk of overfitting. To test for this latter, we ran a set of Monte Carlo simulations as explained in Section\,\ref{section:un}.

We performed inversions using the IRIS data with and without the mm-wavelength point and compared the two. Along with the \ion{Mg}{II} h and k lines we fit two of the UV triplet lines (blended at the IRIS spectral resolution) located between the h and k cores, a selected photospheric line of \ion{Ni}{I} at 2814.4\,\AA~as in \citet{2016ApJ...830L..30D}, but only a few points close to the very core since the line is blended on both wings, and a narrow (0.05\,\AA) window of the only part of the continuum between 2831 and 2834\,\AA~that is relatively line free. All wavelengths are given in air values. 
Whether or not other photospheric lines in the IRIS passband can be included is under investigation.

We did not attempt to fit FUV lines such as \ion{Si}{IV} because recent modeling attempts of these lines with STiC have proven to be quite challenging \citep[see][]{2019A&A...627A.101V}. Therefore, whenever we mention IRIS inversions we refer to inversions of NUV lines and continuum.

In order to fit the selected IRIS NUV spectra alone, we typically only need between seven and nine temperature nodes, but the relatively narrow response function of the mm continua in optical depth coupled with a spatial and temporal dependence requires a more dense node parametrization between $\log\tau\sim[-6,-4.5]$ to ensure that we cover the range of formation heights. We found that 19 temperature nodes (mostly in the chromosphere and upper photosphere) counter-balanced with regularization to impose some degree of smoothness \citep{2019A&A...623A..74D} works best for the majority of the FoV. We can in principle obtain equally good fits with lesser nodes at certain locations (but definitely not in the hottest §\,\ref{Section:shockSignatures} and coolest areas §\,\ref{section:coolChromosphere}), however this requires manual adjustment of their placement which is unfeasible for large datasets. 
We used nine and seven equidistant nodes for $v_{\rm turb}$ and $v_{\rm LOS}$, respectively.

We take into account the IRIS instrumental spectral profile using a Gaussian function with a fullwidth at half maximum of 53\,m\AA~\citep{2014SoPh..289.2733D}. Because all the ALMA frequencies within each sub-band were averaged into one single effective wavelength (1.25\,mm), we could not make use of the full spectral resolution to improve our inversions. However, we do not expect large opacity changes within the 0.1\,mm bandwidth of Band\,6 \citep[see also][]{2019A&A...622A.150J}, and since the mm intensities can be translated to gas temperature via the Planck function which is approximately linear in wavelength within such a narrow wavelength interval, the average Band\,6 brightness temperature essentially corresponds to the brightness temperature of the effective wavelength, with an error that is much smaller than the statistical uncertainties.

We investigated the effect of the IRIS point-spread-function (PSF) on the inversions by first deconvolving the data with a regularized Richardson-Lucy iterative deconvolution algorithm with a measured 1D PSF in the NUV \citep{2018SoPh..293..125C}.  
We found that the difference in the inverted temperatures is not significant, especially in the IRIS+ALMA inversions which are much more dominated by the inclusion of the ALMA data.

% --------------------------------------------- %
\subsection{Response functions}
\label{section:RF}
We computed response functions to temperature perturbations $R_{\rm T}$ as a function of wavelength and logarithmic optical depth numerically to quantify the different sensitivities of the spectral diagnostics to changes in the temperature stratification. This allows us to identify the parts of the atmosphere that are more relevant for the formation of the spectral features under study. Response functions to a temperature perturbation $\Delta T$ are computed by subtracting the (unperturbed) synthetic spectra from the emergent spectra from the perturbed atmosphere for each $\log\tau,$ as summarized in the equation below:
\begin{equation}
R_{\rm T}(\lambda_{\rm i}, \log \tau_{\rm j}) = \frac{I_{\rm syn}(\lambda_{\rm i}, T+\Delta T_{\log \tau_{\rm j}})-I_{\rm syn}(\lambda_{\rm i}, T)}{\Delta T \Delta \log \tau}
,\end{equation}
\noindent where $\Delta \log \tau=0.1$ is the step length in the optical depth grid.

% --------------------------------------------- %
\subsection{Uncertainties}
\label{section:un}

\begin{figure*}[t]
    \centering
    %\sidecaption
    \includegraphics[width=\linewidth]{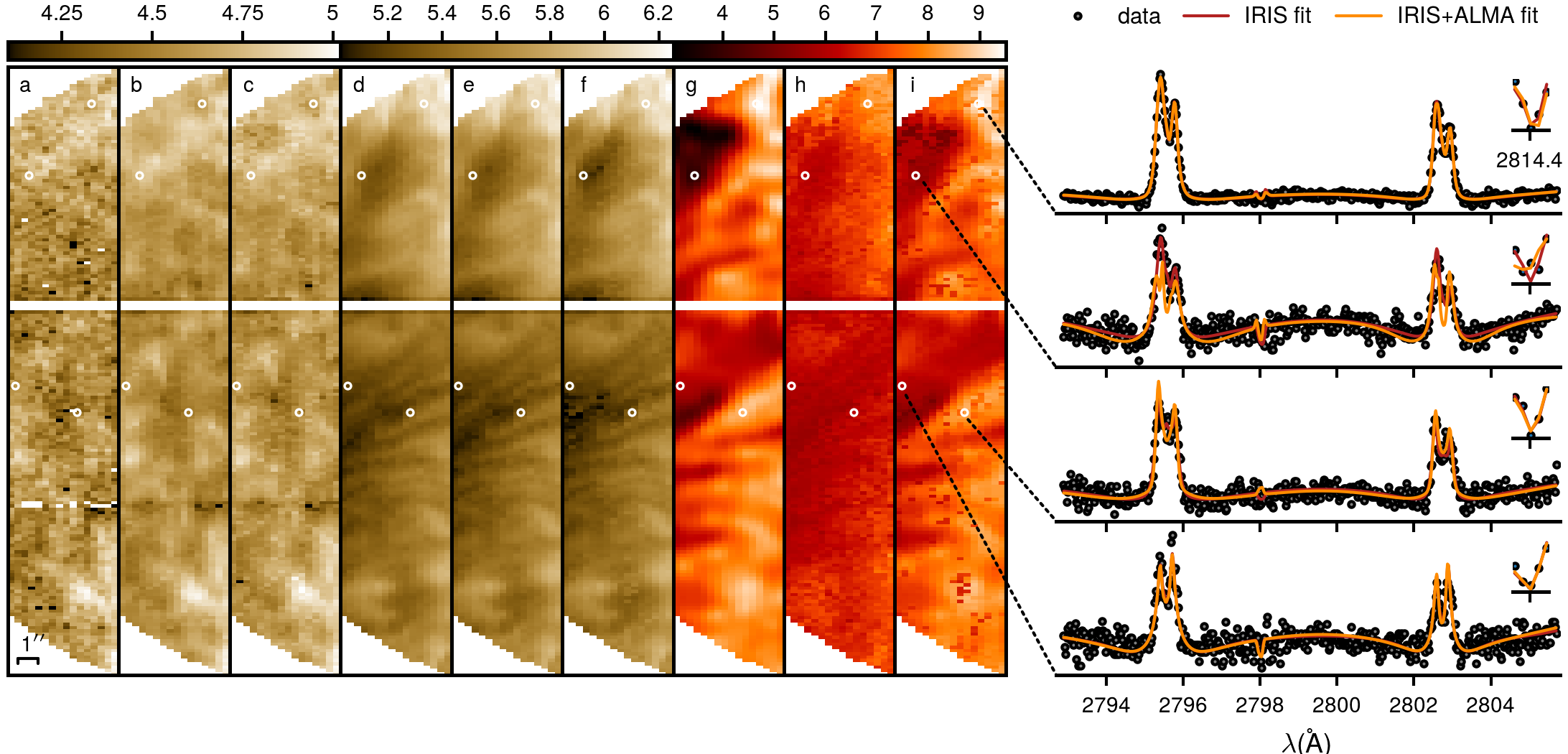} %width=120mm
    \caption{Observed and synthetic IRIS and ALMA spectra. Averaged continuum at 2800($\pm0.1$)\,\AA~({\it panel (a):} observed, {\it panel (b):} synthetic from IRIS fit, and {\it panel (c):} synthetic from IRIS+ALMA fit),  averaged k core wavelengths ({\it panel (d):} observed, {\it panel (e):} synthetic from IRIS fit, and {\it panel (f):} synthetic from IRIS+ALMA fit), and 1.25\,mm continuum ({\it panel (g):} observed, {\it panel (h):} prediction from IRIS fit, and {\it panel (i):} synthetic from IRIS+ALMA fit). The color bars are in units of kiloKelvin. The panels on the right show examples of observed and synthetic IRIS spectra (arbitrary y-scale) at selected locations. The insets show the core of the \ion{Ni}{I} line.}
    \label{fig:fig5}
\end{figure*}

To obtain an estimate of the order of the uncertainties in the inversion parameters, we ran sets of up to $n=100$ Monte Carlo simulations per pixel of synthetic IRIS and ALMA data generated from the best-fit models, adding the derived statistical uncertainties and their respective flux calibration uncertainties (§\,\ref{section:observations}). We also used different, randomly generated model atmospheres as starting guesses in each Monte Carlo simulation. This number of runs was found to provide enough statistics for accurate (to <0.5\%) 68\% confidence intervals on the parameters (temperature, $T$, line-of-sight velocity, $v_{\rm LOS}$, and microturbulence, $v_{\rm turb}$) as a function of $\log\tau$.

We note that this sort of analysis cannot be performed on the entire FoV because it would increase the computational cost by two orders of magnitude, and therefore we are limited to a few representative pixels in the plage and in the quasi-quiet-Sun part of the FoV. However, it is more powerful than estimating uncertainties from response functions alone in the sense that we can account for the expected calibration uncertainties, and test the stability of the solution against different starting guesses. 

In the weakly magnetized parts of the FoV, the uncertainties (1$\sigma$) in the inverted temperatures are expected to be of the order of $\sim$100-200\,K in the photosphere ($\log\tau\sim[-3,0]$), and increasing through the chromosphere up to $\sim$200-600\,K between $\log\tau\sim[-6,-4]$.
The uncertainty in $v_{\rm turb}$ is typically $\lessapprox$1$\rm\,km\,s^{-1}$ at $\log\tau\sim-5$, but increases to $\sim$2$\rm\,km\,s^{-1}$ at $\log\tau\sim-2$ where it is only constrained by the \ion{Ni}{I} line, and to $\sim$4-6$\rm\,km\,s^{-1}$ for $\log\tau\lessapprox-6$ near the formation heights of the h and k cores where the sensitivity is lower. The uncertainty in $v_{\rm LOS}$ is typically $\lessapprox$1$\rm\,km\,s^{-1}$.
The uncertainties are of the same order in the plage region, but here the location and temperature gradient of the transition region seem to be more strongly constrained than in the weakly magnetized areas. Around $\log\tau\sim-5$ the uncertainties are also typically $\sim$50-100\,K lower.
We note that for optical depths much lower than $\log\tau\sim-6$ the results are not trustworthy because none of the diagnostics are very sensitive in this region. 

% ------------------------------------------------------- %
\section{Results}
\subsection{Comparison between synthetic and real data}

Figure\,\ref{fig:fig5} shows a comparison of observed and synthetic images at some averaged wavelengths in the \ion{Mg}{II} spectral window and the 1.25\,mm continuum map. The \ion{Mg}{II} intensities have been converted to brightness temperature for comparison with the ALMA data. No degradation of the IRIS data was done to allow a comparison of the structures at different resolution.
Overall, the ALMA 1.25\,mm image shows  structures that are similar to the \ion{Mg}{II} k (and h) core, but the range and contrast in brightness temperature is quite different: the k line core spans about $\sim$1000\,K in range while the 1.25\,mm shows a wider variation of $\sim$6500\,K from the weakly magnetized areas to the hot plage.

Figure\,\ref{fig:fig5} also shows that, in general, the inversions are able to reproduce both the cores and wings of the k (and h, not displayed) line  very well. However, we find that the inferred parameters from the inversions of IRIS data alone are unable to accurately predict the mm continua (panel (h)). The synthetic image shows traces of the structures that are actually seen in the real data (panel (g)), which gives us confidence that the data alignment is good enough. However,  the brightness contrast is far inferior to the observed, which may be a sign of significant NLTE radiative transfer effects in the formation of the \ion{Mg}{II} lines that hinder accurate inference of chromospheric temperatures. This has been demonstrated for NLTE inversions of the \ion{Ca}{II} 8542\,\AA~line based on simulations \citep{2012A&A...543A..34D}. This is further discussed in Section\,\ref{section:discussion}.

When we combined ALMA with IRIS, we were able to reproduce the observed mm data (panel (i)) much better, while still properly fitting the NUV data in most cases. The best fits are in the weakly magnetic regions with ALMA brightness temperatures in the range $\sim$5000-7000\,K, but also at some locations within the plage region. The more extreme values are more challenging to fit; some of the brightest parts remain underestimated by a few percent, whereas the dark patches are more severely overestimated, meaning that the synthetic 1.25\,mm map still lacks contrast compared to the observed one.
However, the fact that we can reproduce the ALMA data to better than 5\% ($2\sigma$) for the most part in the FoV and that the residuals are centered around zero suggests that the calibration error may not be as large as anticipated, and that there should not be a significant calibration offset. 

\begin{figure*}
    \centering
    %\sidecaption
    \includegraphics[width=\linewidth]{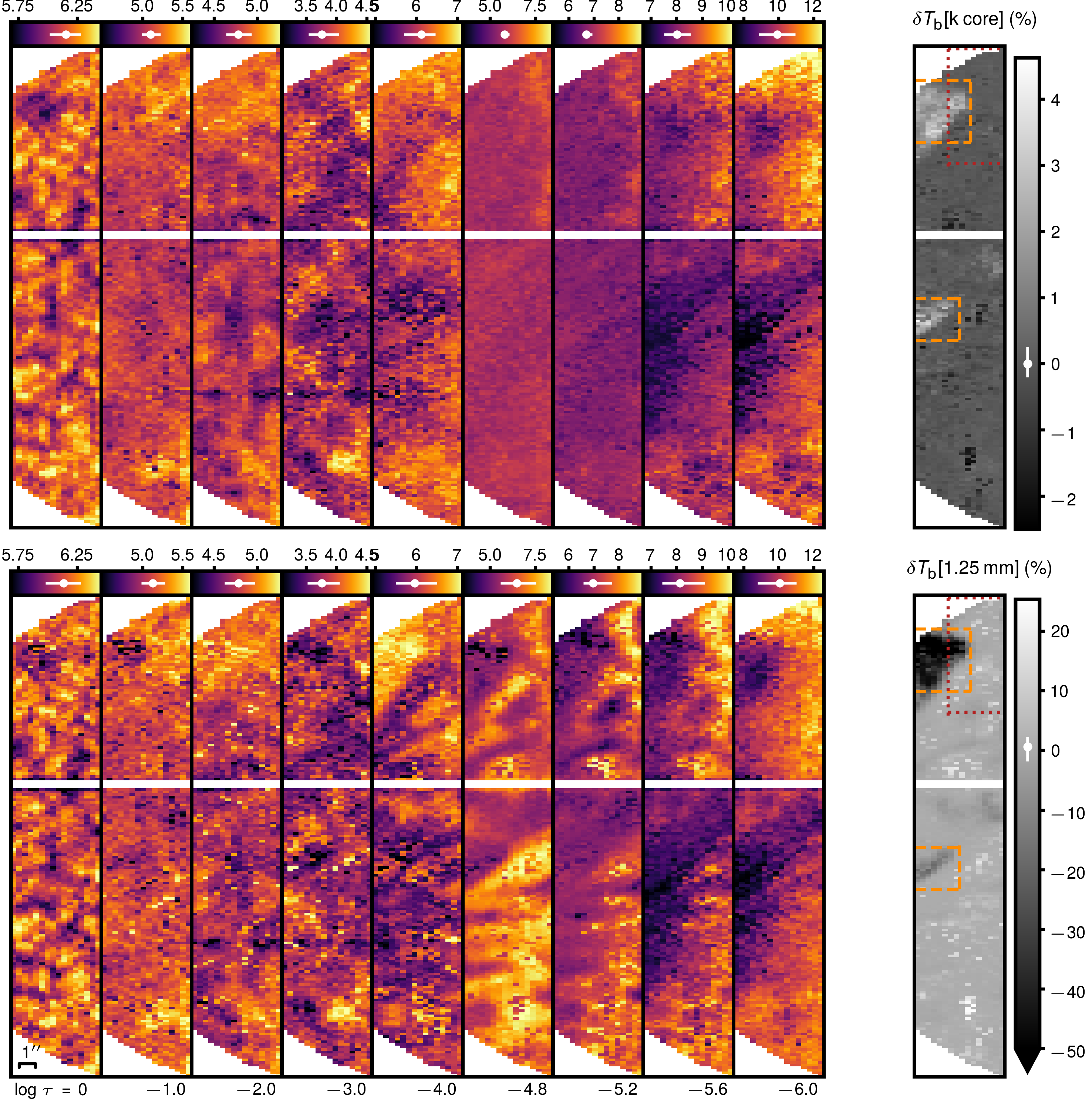}
    \caption{Inferred temperatures from NLTE IRIS and ALMA inversions. Temperature maps (in kK) at different optical depths from IRIS data alone (top left grouped panels) and with ALMA (bottom left grouped panels) corresponding to the synthetic data in Fig. \ref{fig:fig5}. The rightmost panels show the residuals in brightness temperature of the k core (top) and 1.25\,mm (bottom) in the IRIS+ALMA fit. The error bars on the color maps indicate the median and the range between the 16th and 84th percentiles. The regions inside the dashed boxes are further investigated in Fig.\,\ref{fig:Upper}.}
    \label{fig:fig2}
\end{figure*}

The right panels of Fig. 2 show four example IRIS spectra and their best fits at different regions of interest in the FoV. The plage area at the top features the brightest \ion{Mg}{II} profiles in the FoV with (mostly) asymmetric peaks with radiation temperatures of the order of $\sim6100-6400$\,K, which are very well fitted in general. 

The chromospheric hole shows relatively weaker \ion{Mg}{II} h and k emission that is brighter than the mm continua in radiation temperature scale. The inversions of the NUV alone reproduce the line shapes very well, but when adding ALMA to the fit, the code tends to underestimate the \ion{Mg}{II} intensities in order to explain the very low mm brightness. The ALMA image also shows hot elongated structures connecting the plage region to the QS that appear broader than the \ion{Mg}{II} counterparts, which is to some extent due to the lower resolution. The IRIS and IRIS+ALMA fits produce similar synthetic NUV spectra, but whereas the IRIS-only fit puts the UV triplet into absorption, the inversion with ALMA forces it into weak emission. At some locations along these hot streaks the IRIS+ALMA inversions overestimate the h and k peaks by a few percent. The cooler and weakly magnetized areas are generally well fitted in both IRIS and IRIS+ALMA inversions. Here the synthetic UV triplet is found in absorption, and the mm brightness temperatures are very well reproduced (better than 2\%).

% -------------------------------------------------- #
\subsection{Temperature maps: full raster}
\label{results:fullraster}

The example fits in Fig.\,\ref{fig:fig5} show that the synthetic \ion{Mg}{II} profiles in the two inversion modes are similar, but the inferred temperatures must be different so as to explain the ALMA observations. This is emphasized in Fig.\,\ref{fig:fig2} where we show the inverted temperature maps at different optical depths for the IRIS and IRIS+ALMA inversions (same temperature range), and the $T_{\rm b}$ residuals of the k line and the 1.25\,mm continuum in the IRIS+ALMA inversions. 
The inversions show the granulation pattern imprinted on the temperature at $\log\tau=0$, the reversed granulation at $\log\tau=-2$, and an overall differentiation between plage and its surroundings confirming that the plage region corresponds to columns of systematically enhanced temperatures. The standard deviation of the distribution of the residuals at 1.25\,mm is 3\%.

We find the most significant differences between IRIS-only and IRIS+ALMA inversions at around $\log\tau\sim-5$ (and lower depths) where we obtained temperature maps with higher contrast that show much more structure, just like in the experiments with simulated data in \citet{PaperI}. Overall, there is no large difference in the temperature maps at $\log\tau\sim-6$ which means that the 1.25\,mm already has a weak response or none at all in these layers. One exception is during the hot mm transients in plage, which we investigate in more detail in Section\,\ref{section:time_evolving}.

The IRIS+ALMA inversions run into a few problems: (1) the IRIS+ALMA inversions show more "salt and pepper" noise as a result of more troublesome convergence at selected locations; (2) the hot pixels around $\log\tau\sim-4$ with temperatures approaching $\sim7000$\,K are mostly artifacts resulting from the overestimation\footnote{This is not an exclusive problem of the IRIS+ALMA inversions though, and it could be related to the node parametrization.} of the UV triplet emission, and (3) the synthetic cool patches are not as cool as in the ALMA data, and not as bright as in the IRIS data. 

Running the two inversion schemes (IRIS and IRIS+ALMA) with the same number of nodes and regularization weights in all parameters (§\,\ref{section:setup}) leads to good fits for most pixels in the FoV. However, in the coolest ALMA pixels, that approach fails to reproduce the ALMA brightness temperatures and returns an error as large as 100\%.
% ---------------------------------------------------- %
\subsection{Chromospheric holes}
\label{section:CH}

\begin{figure*}
    \centering
    \includegraphics[width=\linewidth]{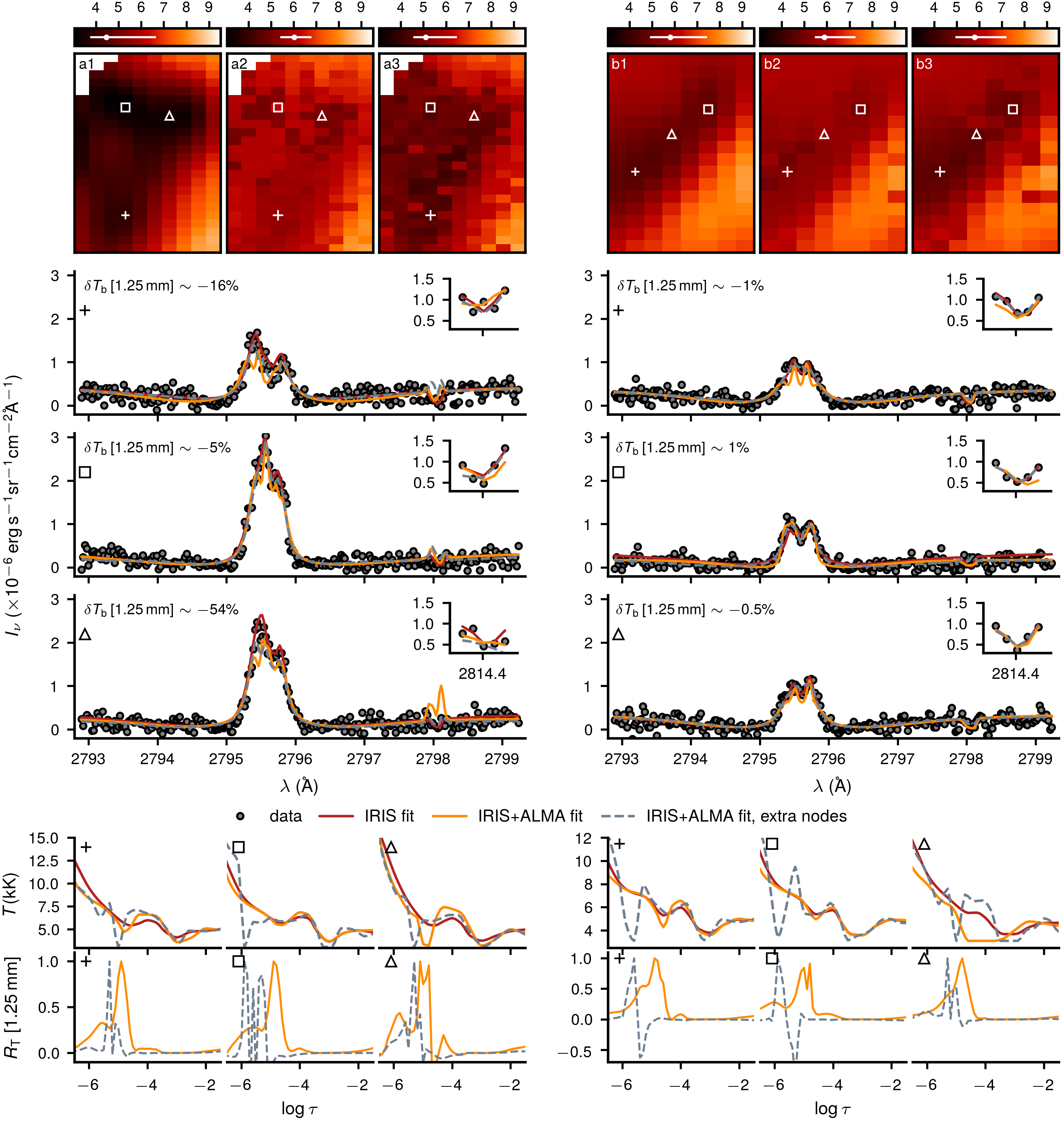}
    \caption{Chromospheric holes. Observed ALMA 1.25\,mm brightness temperature ({\it panels (a1) and (b1)}) and synthetic images corresponding to the IRIS+ALMA fit in Fig.\,\ref{fig:fig2} ({\it panels (a2)\ and (b2)}), and with an increased number of nodes ({\it panels (a3) and (b3)}) for the two regions enclosed by the dashed boxes in Fig.\,\ref{fig:fig2}. The errorbars on the colormaps indicate the median and the range between the 16th and 84th percentiles. We show part of the observed and synthetic IRIS spectra and the respective inverted gas temperatures and normalized response functions (1.25\,mm) at three selected locations as indicated by the different markers for both regions ($a$ and $b$).}
    \label{fig:Upper}
\end{figure*}

Figure\,\ref{fig:Upper} shows the regions enclosed by the dashed boxes in Fig.\,\ref{fig:fig2} where we find the largest misfits to the mm continua. 
The fact that the dark patches in the ALMA data are not visible in the 2800\,\AA~continuum or in the \ion{Ni}{I} line core, or even in the Hinode \ion{Fe}{I}\,6302 cores (not displayed), suggests that these "holes" are relatively localized in height and do not extend much further down to the upper photosphere. The fact that they are also not particularly dark in the cores of the h and k lines is partly due to the IRIS PSF (see Fig.\,\ref{fig:overview}). This was investigated by first deconvolving the data with the instrumental profile, and the result is an increased intensity contrast that nevertheless has only a small impact on the inverted atmospheres. More important are the 3D radiative transfer effects: scattered radiation fills the chromospheric hole raising the local source function. This poses a limitation on our inversions. 

The first inversion results suggested the presence of localized depressions in the temperature stratification in the chromosphere at these locations. For this reason, we decided to experiment with a large number of nodes (up to 30) in temperature to allow for additional degrees of freedom. The effect is counteracted with an increase in the regularization weight to reduce the oscillatory behaviour.

We find that more complicated temperature profiles are needed to simultaneously reproduce the cool 1.25\,mm continua and the \ion{Mg}{II} emission lines at the chromospheric holes. The code manages to lower the mm intensities by placing a "hole" in temperature higher up in the atmosphere in the range $\log\tau\sim[-6,-5]$ where the response functions of the 1.25\,mm continuum are sharply peaked. This feature does not occur in the IRIS-only inversions.
However, overly increasing the number of nodes not only increases the computational cost but could also lead to overfitting that may not be entirely mitigated by the regularization term, which in turn leads to more inversion noise. Doing this kind of inversion with many nodes for the entire FoV has been shown to lead to poor results in previous experiments, and is nevertheless unjustified because we obtain good fits with simpler models. In any case, it may be that these columns require a more complicated temperature stratification that it is more difficult to parameterize \citep[see e.g., ][]{2012A&A...543A..34D}. 

\begin{figure}
    \centering
    \includegraphics[width=\linewidth]{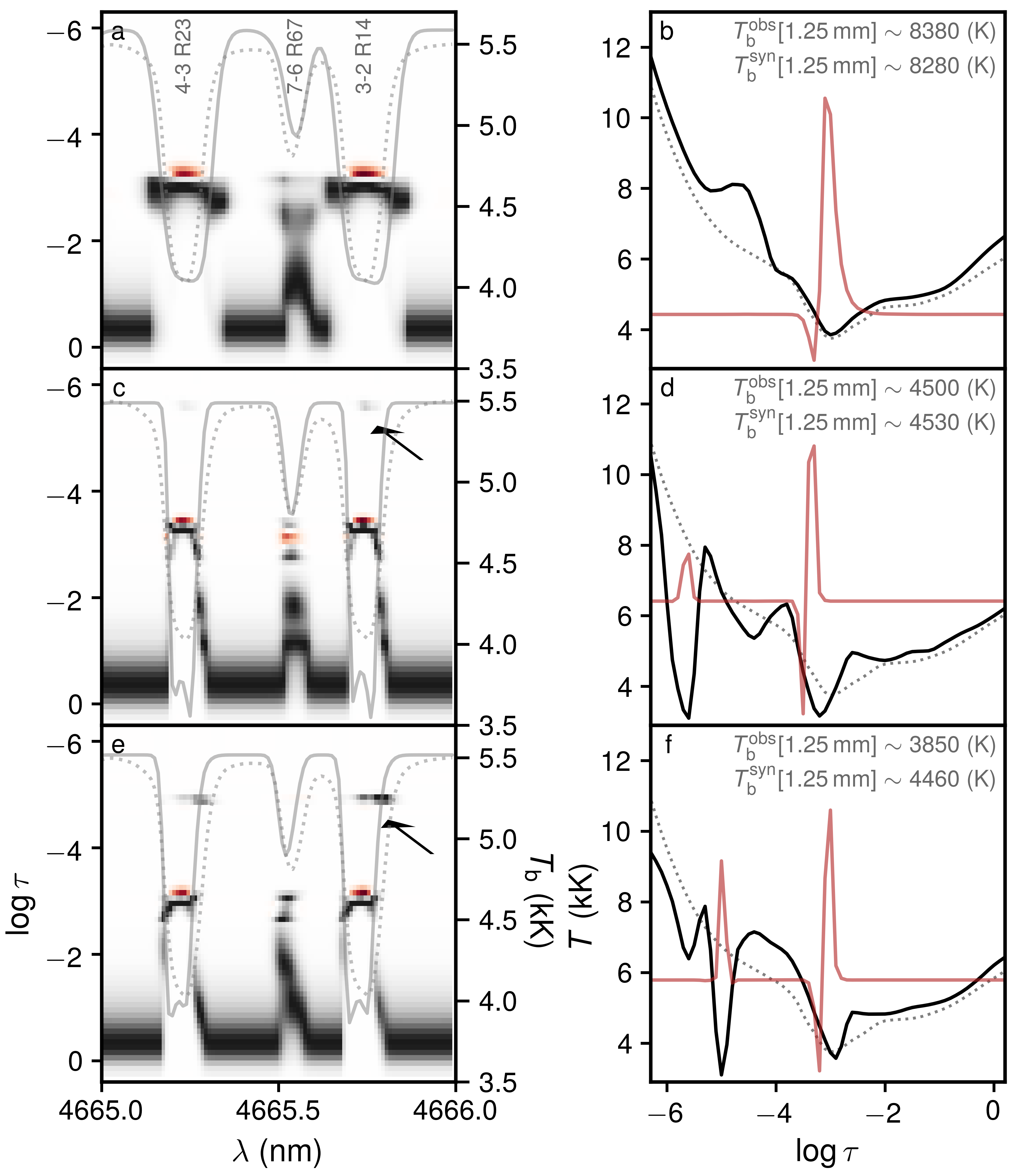}
    \caption{ Fundamental vibration rotation CO lines at 4.67\,$\rm \mu m$ and response functions. {\it Left panels:} Synthetic CO lines (solid lines) and FoV average (dotted lines) in brightness temperature scale, and response functions to temperature as a function of optical depth in plage (a) and in the cold pocket (c, e); {\it Right panels:} Corresponding temperature profiles (solid lines, FoV average (dotted lines), and response functions to temperature at the core of the CO 3-2\,R14 line in plage (b) and in the cold pockets (d, f). The observed and synthetic ALMA brightness is indicated. }
    \label{fig:CO}
\end{figure}

Low temperatures ($\lesssim4000$\,K) occurring at a few hundred kilometers above the temperature minimum have previously been inferred from observations of the molecular gas component, namely CO, in the QS \citep[e.g.,][]{1994Sci...263...64S,1994ApJ...432L..67U,1996ApJ...460.1042A}.
CO forming at even greater heights was deemed unlikely because the UV diagnostics implied a generally hot chromosphere, and 1D nonequilibrium chemistry calculations constrain the formation height of CO to no greater than $z\sim700$\,km \citep{2003ApJ...588L..61A}. 
Cool gas at chromospheric heights has been reproduced in simulations, such as the hydrodynamical models of \citet{Carlsson95, Carlsson97}, and the 2D and 3D MHD simulations of \citet{2011A&A...530A.124L,2012A&A...543A..34D,2016A&A...585A...4C,2017ApJ...847...36M} to name a few. The 2D and 3D simulations indeed show that while the bulk of CO is formed in the photosphere, CO concentrations can remain high in the cool regions of the higher layers except when there are shocks. 

This led us to the conclusion that perhaps the cool ($\lesssim 4000$\,K) temperatures observed with ALMA Band\,6, which is expected to probe a range of heights typically within $\sim700-1200$\,km \citep[e.g.,][]{2007A&A...471..977W,2015A&A...575A..15L, PaperI}, may allow for the cool CO pockets (or clouds) to form. 
We investigated the problem of CO formation in the chromosphere using our models constrained by IRIS and ALMA observations.
We computed a few vibration-rotation CO lines in the commonly observed near-infrared (NIR) range around 4.67\,$\rm \mu m$ assuming LTE. 

Figure\,\ref{fig:CO} shows some synthetic vibration-rotation CO lines along with response functions to temperature perturbations at selected locations in the cool (dark) patches where the ALMA misfit is smaller than 15\%. Even in the plage region where the overall temperature stratification is enhanced compared to the weakly magnetized areas, the presence of a temperature minimum of $\sim4000(\pm100)$\,K allows CO to form at those heights, but the much hotter chromospheric temperatures impede its formation in higher layers.
At some locations within the chromospheric hole the temperature minimum is placed below 4000\,K which causes the lines to go deeper (down to $\sim$3500\,K), but we also find a small contribution of the chromospheric hole in the range $\log\tau\sim[-6,-5]$ to the CO line profiles, which is an indication that CO could form in the cool chromosphere. Time-dependent modeling would be needed for more accurate modeling of CO in chromospheric conditions \citep[e.g.,][]{2003ApJ...588L..61A,Wedemeyer2005}.

% ---------------------------------------------------- %
\subsection{Microturbulence: full raster}
\label{section:microturbulence}
\begin{figure}
    \centering
    %\sidecaption
    \includegraphics[width=0.75\linewidth]{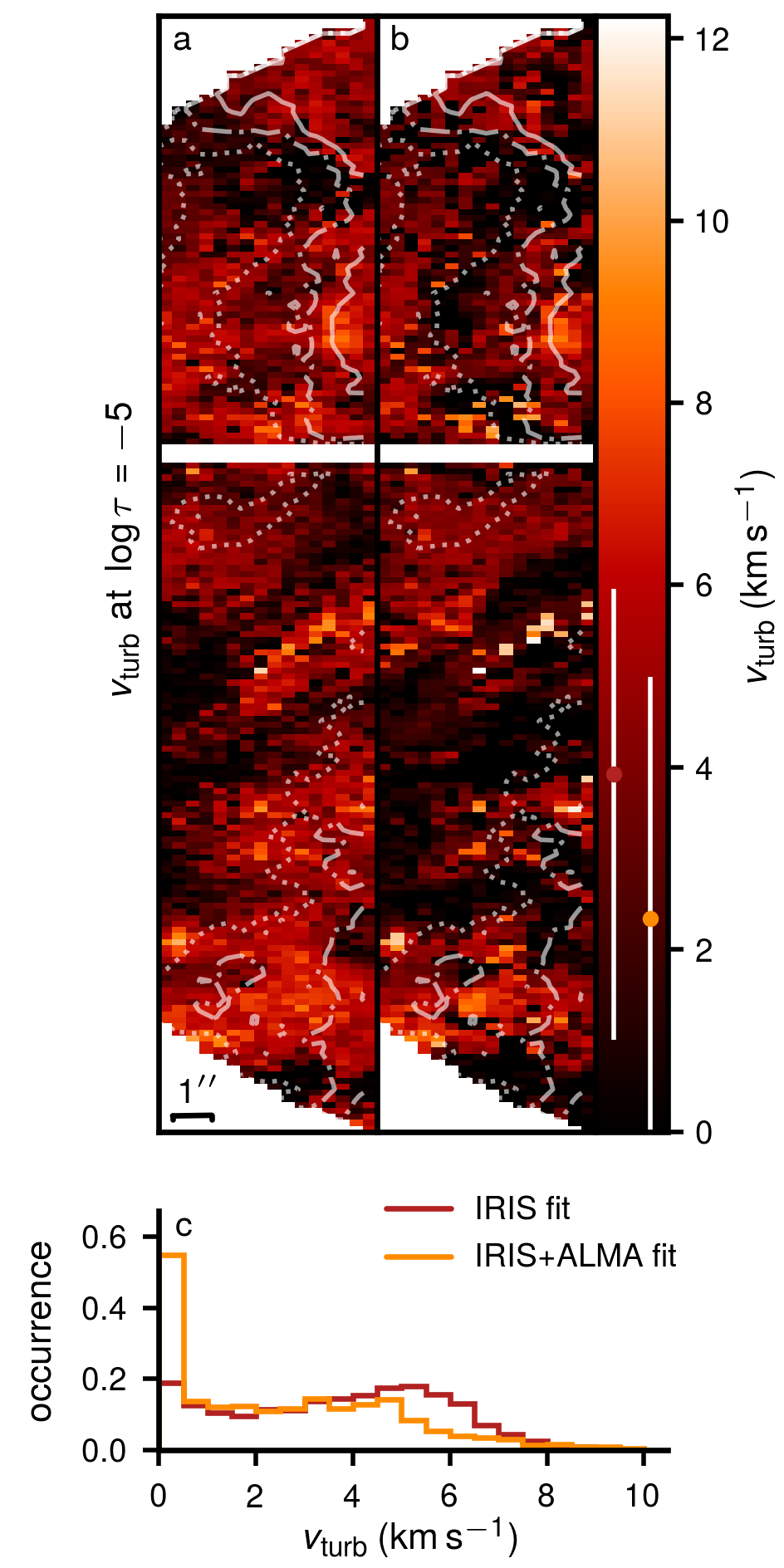}
    \caption{Microturbulence distributions. \textit{Panels (a)}~and \textit{(b)}: Microturbulence maps at $\log\tau=-5.0$ for the IRIS and IRIS+ALMA fits, respectively; the contours are brightness temperature thresholds (5600\,K - dotted, 5800\,K - dash-dotted, and 6000\,K - solid) of the average of the k2v and k2r peaks; the error bars in the color bar indicate the median and the range between the 16th and 84th percentiles. {\it Panel (c)}: Corresponding (normalized) histograms.}
    \label{fig:vturb}
\end{figure}

In broad terms, we find that the microturbulence is generally higher towards the plage region and increases with height. There are differences between the IRIS and IRIS+ALMA inversions as shown in Fig.\,\ref{fig:vturb} where we compare the distribution of microturbulence in the IRIS and IRIS+ALMA fits at $\log\tau=-5$, which is the average optical depth at which the response function of the 1.25\,mm continuum to temperature  peaks.

The median values for the entire FoV are approximately $\sim4\,\rm km\,s^{-1}$ and $\sim2\,\rm km\,s^{-1}$ for the IRIS and IRIS+ALMA inversions respectively, which shows that ALMA is helping to constrain temperatures (and electron densities) and reduce the need for larger turbulent broadening. 
The IRIS+ALMA inversions have a more skewed $v_{\rm turb}$ distribution towards zero at $\log\tau=-5.0$, and a fraction of the pixels have this parameter effectively reduced to near zero. In the weakly magnetized areas we find similar average values for both schemes $\sim3(\pm1)\,\rm km\,s^{-1}$. For the hottest part of the plage in this raster (approximately defined as the area inside the $T_{\rm b}=6000$\,K contours of the k line), we find $\sim5(\pm)2\rm\,km\,s^{-1}$ and $\sim3(\pm)2\rm\,km\,s^{-1}$ for the IRIS and IRIS+ALMA inversions, respectively. We note that there is a time dependence on the values of microturbulence in plage (§\,\ref{section:time_evolving}).

The STiC code tries to raise the temperature in the chromosphere and to move the location of the transition region in column mass, but it can only do so to a certain extent while imposing a smooth atmosphere and without overestimating the emission in the subordinate triplet lines, which is rarely seen. Therefore, STiC is forced to convert the remaining excess line broadening into large microturbulences in the line forming region. 
Some parts of the FoV, particularly towards the plage area, still require large values of microturbulence (up to $\sim8\,\rm km\,s^{-1}$). This may be partly because the mm continuum is still not perfectly fitted there and is typically underestimated by a few percent. It may also confirm the significant nonthermal broadening within the plage regions as suggested by the widths of the \ion{O}{I} line in the FUV  \citep{Carlsson_2015}. Indeed, we ran an inversion test imposing $v_{\rm turb}=0$ and found that, especially in plage, it is difficult to obtain a satisfactory fit to the \ion{Mg}{II} lines even when the 1.25\,mm is fitted at the same time.

We note that it is possible that the formation height of the mm continuum is significantly shifted in optical depth such that we would not see its effect at exactly $\log\tau=-5.0$. We further investigate the effect of ALMA on the inferred microturbulence velocities in Section\,\ref{section:time_evolving}.

% ------------------------------------- %
\subsection{Time evolution of the plage region}
\label{section:time_evolving}

\begin{figure*}
    \centering
    \includegraphics[width=\linewidth]{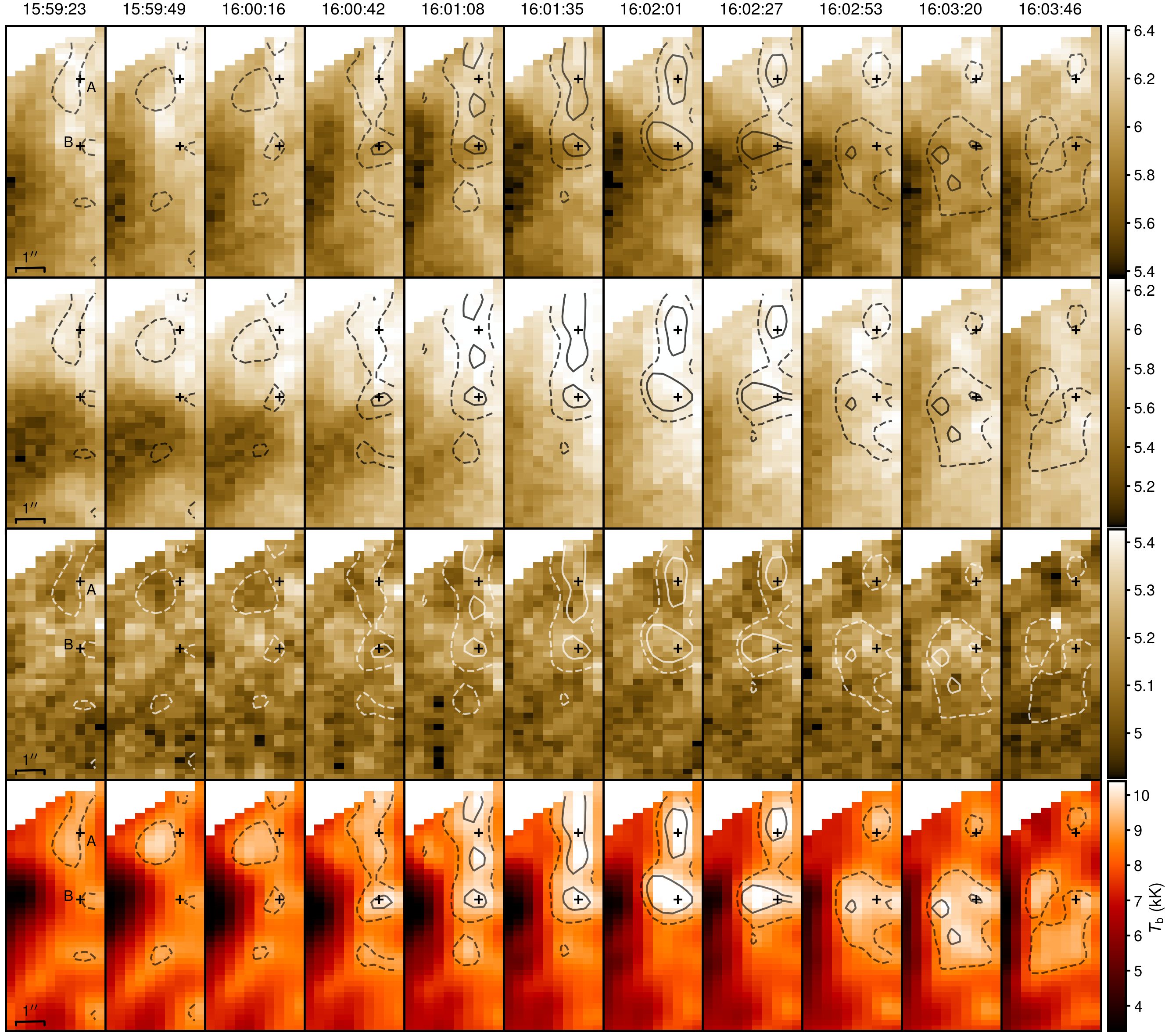}
    \caption{Time evolution of ALMA brightenings. From top to bottom: k2v, k2r, \ion{Ni}{I} core and 1.25\,mm continua in brightness temperature scale evolving in time from the left to the right. We superimpose the $T_{\rm b}[\rm 1.25\,mm]$ contour levels of 9000\,K (dashed) and 10000\,K (solid) in all panels.}
    \label{fig:bursts}
\end{figure*}

Now we focus on the time evolution of a subfield located in the plage region as indicated by the dotted box in Fig.\,\ref{fig:fig2} to investigate whether there are significant variations of the parameters with time. We inverted 18 time steps, which is about $\sim7.5$\,min (before ALMA pauses for calibration), using either IRIS data alone or in combination with ALMA, and compared the two. 

Figure\,\ref{fig:bursts} displays the evolution of the subfield as seen in the core of the k line, \ion{the Ni}{I} core, and ALMA 1.25\,mm for the first 11 time steps. The time stamps correspond to the center of the IRIS raster. 
Initially, there is a region of enhanced emission in the blue peak of the k line (k2v) around location A with $T_{\rm b}\sim6400$\,K that is cospatial with a region of $\sim$9000\,K in the ALMA map. In the minutes that follow there is an enhancement of the red peak (k2r) intensity in an region that is a few arcseconds across where we also see brightenings in the mm continuum reaching over 10\,000\,K. Such high temperatures only last for $\sim$2\,min, but a more extended area with $T_{\rm b}\sim9000$\,K is present after this event. 

The evolution of the bright mm features is not echoed in the UV triplet maps or it happens below the noise level. We see hints of enhanced emission towards the plage region in the core of the \ion{Ni}{I} line, which suggest that the upper photosphere is hotter than average at these locations, but the contrast is poor. 
The brightenings occur in the vicinity of a small pore, in an area where the magnetic field is predominantly unipolar, at least at the 1\arcsec~resolution of the HMI magnetograms. We do not find any obvious counterpart in the SDO/AIA images apart from the already enhanced average plage emission compared to the QS. Therefore, we conclude that these short-lived hot features are also probably purely chromospheric phenomena, and they are different from Ellerman bombs and UV-bursts that would shine well above the average 1600\,\AA~and 1700\,\AA~continua \citep[e.g.,][and references therein]{2018SSRv..214..120Y,2019A&A...626A...4V}. 

\begin{figure*}
    \centering
    \includegraphics[width=\linewidth]{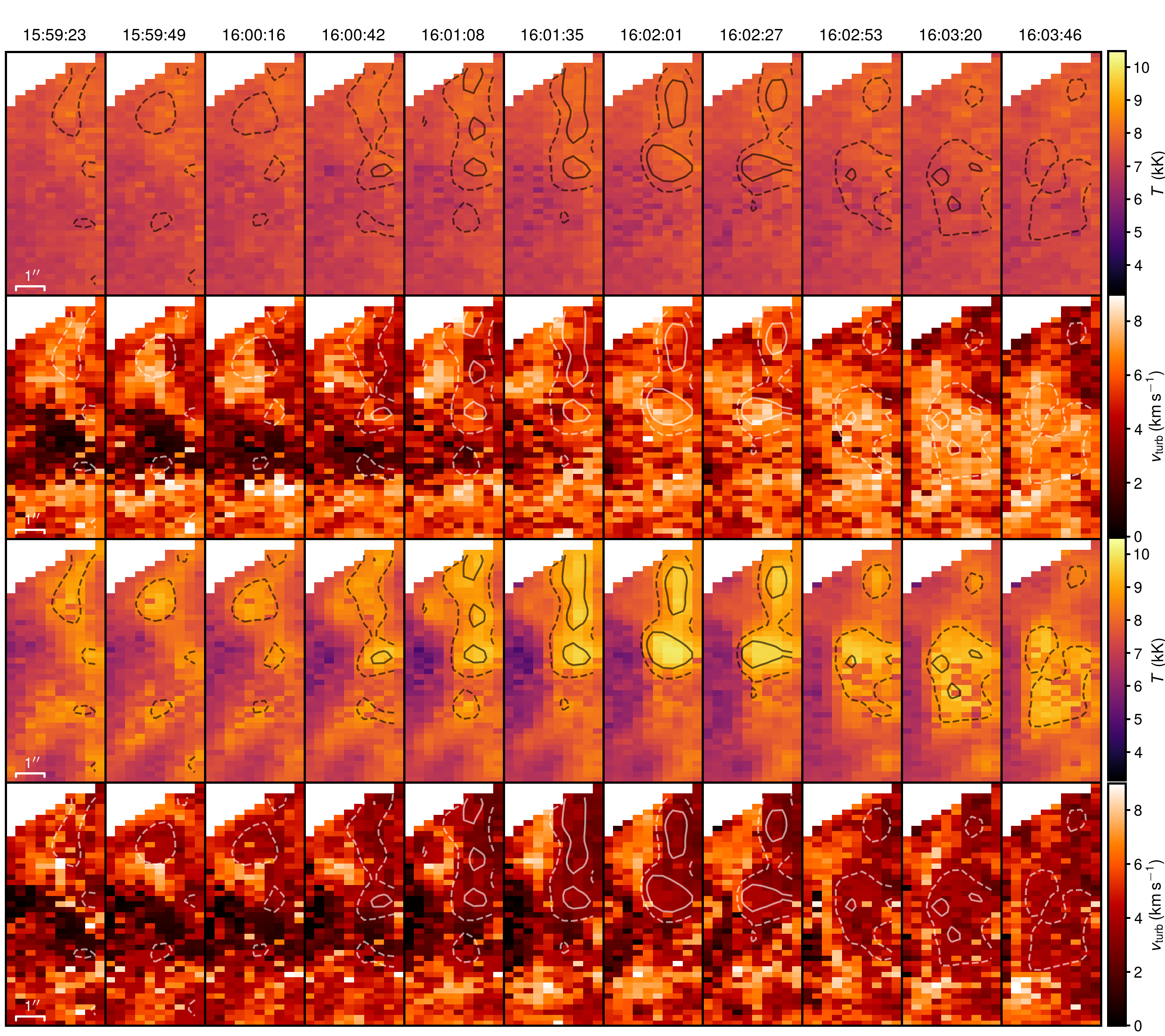}
    \caption{Time evolution of the ALMA brightenings. The panels show temperature and microturbulence averaged over the range ($\log\tau=[-5.8, -4.7]$) inferred from IRIS NUV spectra alone (top two rows), and from IRIS+ALMA inversions (bottom two rows) evolving in time from the left to the right. We superimpose the $T_{\rm b}[\rm 1.25\,mm]$ contour levels of 9000\,K (dashed) and 10000\,K (solid) in all panels. The range in the color bars is the same in two inversion schemes to emphasize the differences in contrast.}
    \label{fig:bursts2}
\end{figure*}

\begin{figure}
    \centering
    \includegraphics[width=\linewidth]{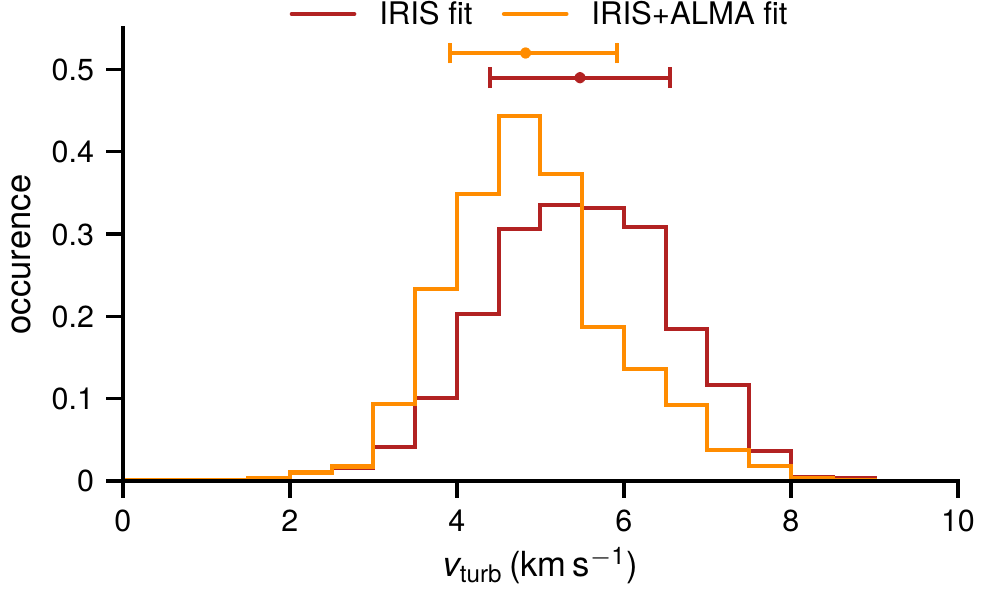}
    \caption{Averaged distributions of microturbulence in plage. The (normalized) histograms correspond to depth-averaged microturbulences for all inverted time steps within the plage region (see text) for the two inversion schemes. The horizontal bars indicate the median 16$^{\rm th}$ and 84$^{\rm th}$ percentiles.}
    \label{fig:vturbdists}
\end{figure}

Figure\,\ref{fig:bursts2} shows the time evolution of the inverted temperatures and microturbulence from IRIS and IRIS+ALMA inversions, both of which were averaged within the optical depths where the 1.25\,mm continuum responds to temperature perturbations.
We find significant differences in the two parameters obtained from the two schemes, in line with the results obtained for the full raster (§\,\ref{results:fullraster}), which emphasizes that ALMA has a systematic, meaningful effect on the inversions. 
The evolution of the NUV and mm diagnostics is well reproduced by the inversions. The residual in the ALMA data is typically better than 2\% within the 9000\,K contours, and about 5\% on average in the rest of the subfield.

The inverted temperatures from IRIS data show enhanced chromospheric temperatures ($\sim7000-7500$\,K in the range $\log\tau\sim[-6,-5]$) in the plage region that match the location of the hot ALMA brightness temperature contours to some extent, but they definitely do not reproduce the rapid temperature enhancements that the ALMA data displays. Instead, STiC transforms the slight excess of line broadening and peak separation into relatively large microturbulence (up to $\sim9\rm\,km\,s^{-1}$) that increases throughout the event within the hot regions in the ALMA data. This suggests that the source functions of the \ion{Mg}{II} lines do not sense the temperature changes at these locations or times.
The picture drawn by the combined inversions of IRIS and ALMA data is different: the chromosphere features much hotter ($\sim9000-11\,000$\,K) regions, which are needed to reproduce the mm continuum approximately within the same range of optical depths ($\log\tau\sim[-6,-5]$) but with a clear shift to lower depths. 

The depth-averaged microturbulence distributions in the magnetic plage ($B_{\rm LOS}>100$\,G in the HMI magnetogram) for all inverted time steps are plotted in Fig.\,\ref{fig:vturbdists} for both inversion schemes. 
The values were averaged over the range $\log\tau=[-6,-4]$, which are the layers most relevant for the formation of the h and k line cores and inner wings according to our models, for comparison with the nonthermal widths of the \ion{O}{I} line in the literature. 
We find that inversions with ALMA tend to favor smaller values in the plage region for different time steps, but the average values of the distributions are not particularly distinct. 
The median  16$^{\rm th}$ and 84$^{\rm th}$ percentiles of the distributions are $5.5^{6.6}_{4.4}\rm\,km\,s^{-1}$ and $4.8^{5.9}_{3.9}\rm\,km\,s^{-1}$ in the IRIS and IRIS+ALMA inversions, respectively, and the uncertainty is typically $\sim1-2\,\rm km\,s^{-1}$.
We note that in the inversions with ALMA the areas where $v_{\rm turb}>6\rm\,km\,s^{-1}$ correspond to pixels where the mm continuum is most underestimated ($\sim5\%$).

\begin{figure*}
    \centering
    \includegraphics[width=\linewidth]{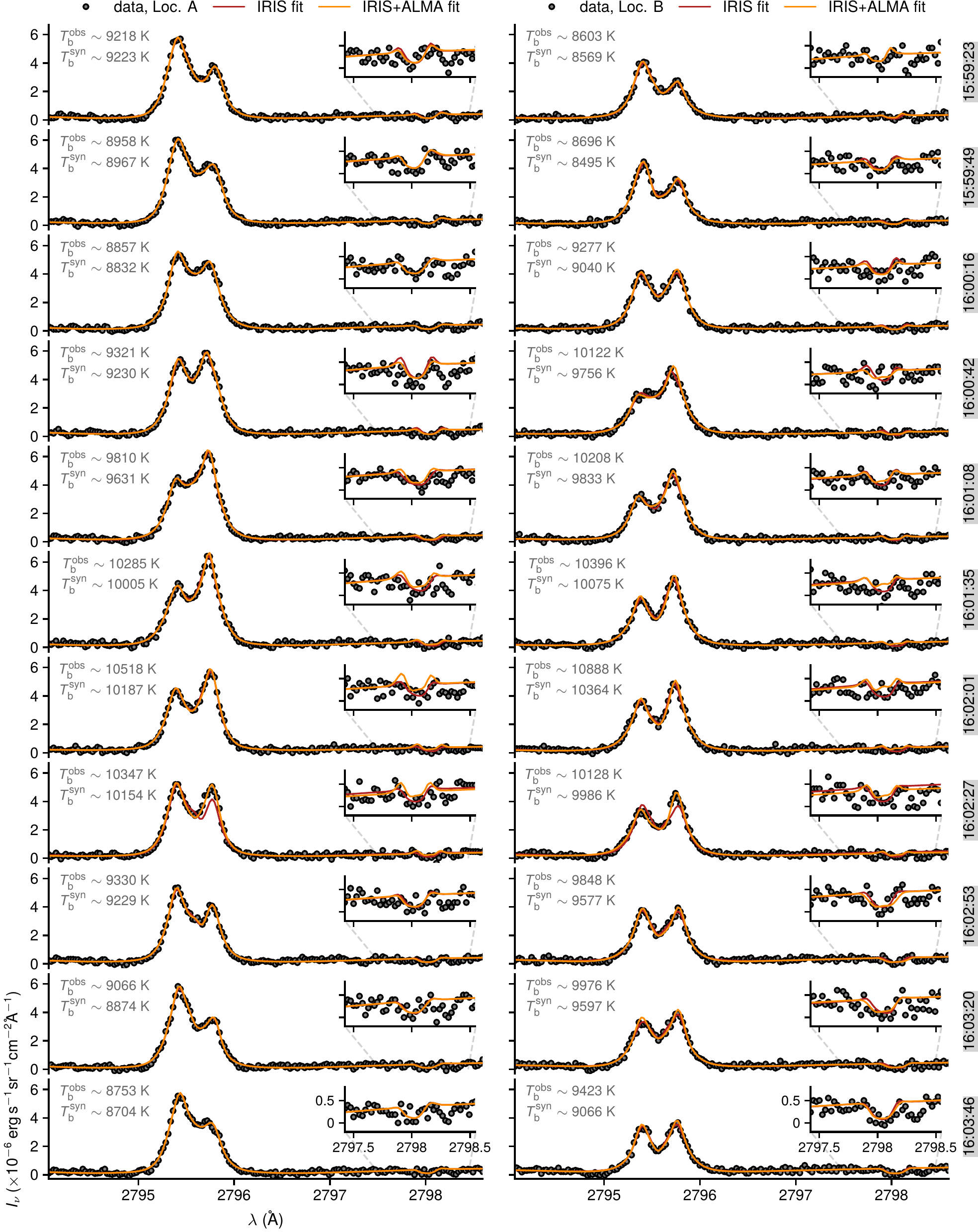}
    \caption{Time evolution of the IRIS NUV spectrum in the hot pockets. The observed spectra and best fits in the range around the k line corresponding to the pixels A and B marked in Fig.\,\ref{fig:bursts}. Spectra are ordered by time stamp from the top to the bottom. The insets zoom into the region around the UV triplet blend. The brightness temperatures of the observed and synthetic 1.25\,mm continua  are indicated.}
    \label{fig:Spectra evolution}
\end{figure*}

Figure\,\ref{fig:Spectra evolution} shows in more detail the time evolution of the spectra at the two selected locations marked in Fig.\,\ref{fig:bursts} in the FoV within the ALMA brightest areas for the first 11 time steps. The \ion{Mg}{II} lines are broad and double-peaked with a peak separation of about 35\,$\rm km\,s^{-1}$ on average. The 1.25\,mm brightness temperature ranges from $\sim$8600\,K to $\sim$10900\,K and increases throughout the event. 
The average brightness of the h and k peaks remains approximately constant during that time span, but there is a (periodic) shift in the peak asymmetry from stronger k2v (blue peak) to k2r (red peak) emission and a broadening of the profiles that seems to coincide with the onset of the mm continuum enhancement. All of these features are very well reproduced by STiC, and the fits for the two inversion schemes (IRIS and IRIS+ALMA) are almost indistinguishable.

\begin{figure}
    \centering
    \includegraphics[width=\linewidth]{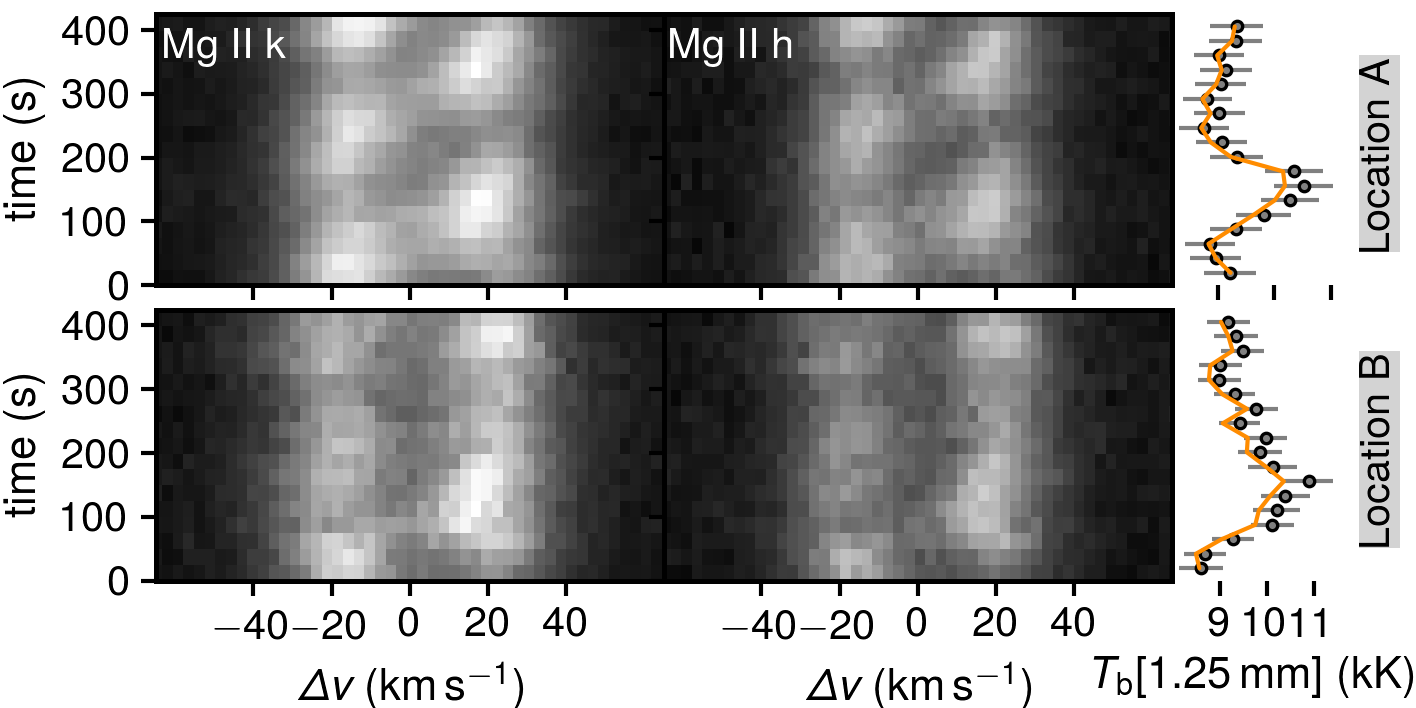}
    \caption{Oscillations in the \ion{Mg}{II} lines. Wavelength--time diagrams for the locations A and B marked in Fig.\,\ref{fig:bursts}. The panels on the right show the variation of the observed (dots) and synthetic (solid line) ALMA brightness temperature.}
    \label{fig:shocks}
\end{figure}

The seemingly periodic nature of this effect is more clearly seen in the wavelength--time diagrams in Fig.\,\ref{fig:shocks} for the full inverted time span. The cadence is 25\,s. The sensitivity to velocities from IRIS observations alone provides key information that allows us to interpret the ALMA brightenings.
At location A (and neighboring pixels) we find a shock wave signature in the \ion{Mg}{II} peak intensities with periods between 3.5 and 4\,min. 
The same trend is not so obvious at location B, but this latter nonetheless shows evidence of outflows that occur when the mm brightness is increased. The modulation of the peak intensities of the \ion{Mg}{II} lines is even less clear (if present at all) in the weakly magnetized areas of the FoV.

\begin{figure}
    \centering
    \includegraphics[width=\linewidth]{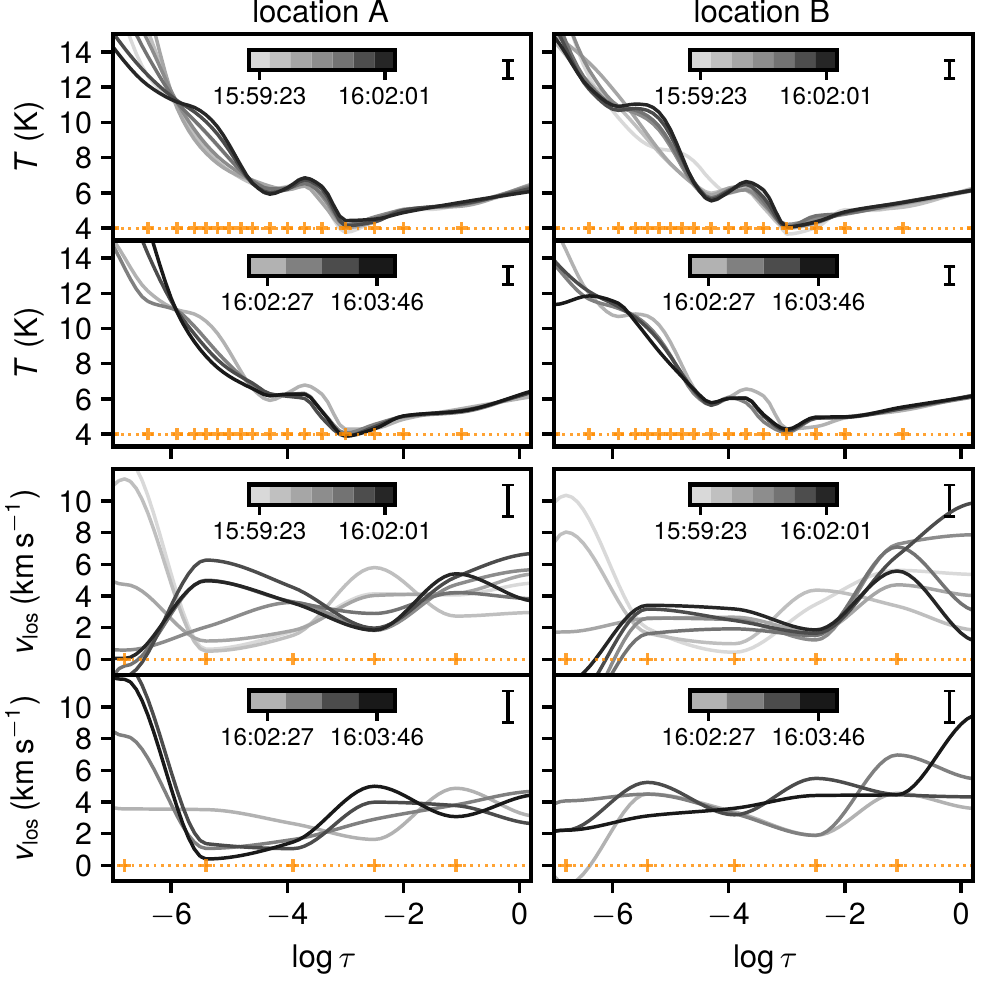}
    \caption{Time evolution of the atmosphere in the plage brightenings. The inferred temperature and velocity along the line-of-sight corresponding to the fits shown in Fig.\,\ref{fig:Spectra evolution}. We subdivide each panel into two phases, namely the temperature rise and cooling, and color-code their progression in grayscale as shown by the color bars. The typical error bars and the node locations are indicated by the vertical bars and crosses, respectively.}
    \label{fig:locAB_evolution}
\end{figure}

The main inversion results at these two locations (Fig.\,\ref{fig:locAB_evolution}) are a $\sim$2000\,K increase in temperature around $\log\tau\sim-5.6$, a steepening of the temperature gradient above the temperature minimum, and a change in the sign of the velocity gradient in the range $\log\tau\sim[-7,-5.5]$, where the cores of the \ion{Mg}{II} h and k lines are sensitive to velocity perturbations in plage, from positive to negative along the LOS in just under 2\,min. The latter can be interpreted as a transition from downflows to upflows. 

Both changes in temperature and velocity gradient are well above the estimated uncertainties. We note that the error bars in Fig.\,\ref{fig:locAB_evolution} correspond to depth-averaged (1$\sigma$) uncertainties, and these can be lower in the upper photosphere than in the upper chromosphere (§\,\ref{section:un}). 
 
The changes in $v_{\rm LOS}$ in the higher chromosphere are replicated in deeper layers between $\log\tau\sim[-3,-2]$ which are set by the k1v/k1r and h1r/h1r features from above, and the \ion{Ni}{I} line from below.
We note that the velocity node at $\log\tau\sim-1$ is poorly constrained by our data, that is, the combined sensitivity range of the \ion{Mg}{II} and \ion{Ni}{I} lines lies mostly within $\log\tau\sim[-7,-2]$. The relationship of these brightenings to shocks is discussed in Section\,\ref{Section:shockSignatures}.

% ------------------------------------- %
\section{Discussion}
\label{section:discussion}

We investigated the physical properties of the solar atmosphere at the edge of a plage region as observed by IRIS and ALMA. Plage are hotter and have higher electron densities than the QS. Both effects contribute to raising the observed brightness temperatures in the mm range via the linear dependence of the source function on temperature, and the increase of thermal bremsstrahlung opacity that raises the observed mm continuum to higher chromospheric layers which are also hotter. 
The ALMA data also feature much cooler ($\sim3000$\,K) temperatures in the weakly magnetic areas which shows the multi-thermal character of the chromosphere and the importance of ALMA as a chromospheric diagnostic.

One of the most important findings is that the synthetic mm spectra predicted from the model chromospheres inferred from \ion{Mg}{II} lines alone do not accurately reproduce the observed high contrast and range in brightness temperature of real ALMA data.
Apart from the ALMA flux calibration uncertainty (§\,\ref{section:observations}), this could be partly due to: (1) NLTE and 3D radiative transfer effects in the \ion{Mg}{II} line cores that make them only partially coupled to the local temperature because the source function is more dominated by photon scattering than collisions, and (2) nonequilibrium ionization affecting the free-free opacity of the mm continuum. 

Even in the presence of time-dependent hydrogen ionization, inversions with ALMA assuming statistical equilibrium are expected to improve the inferred depth-averaged temperatures because they help to set the chromospheric gradient \citep{PaperI}. 
The inclusion of other chromospheric lines such as the H and K lines of \ion{Ca}{II} would likely improve the inference \citep{PaperI}, and therefore it would be interesting to investigate whether predictions of the mm continuum are improved when visible and UV diagnostics are combined.

In broad terms, our results confirm that ALMA helps to constrain temperatures (and electron densities) in inversions of (single-atom) NLTE lines, and that it may help to reduce the need for such large (up to $\sim8\rm\,km\,s^{-1}$) microturbulence broadening in the \ion{Mg}{II} resonance lines at certain locations. Especially in the very cool (§\,\ref{section:coolChromosphere}) and very hot regions (§\,\ref{Section:shockSignatures}), the IRIS+ALMA inversions infer a temperature and density profiles for $\log\tau\lesssim-4$ that is significantly different from the ones obtained by the NUV data alone. However, we also found cases where the inverted atmospheres are very similar. Overall, the contrast in temperature increases and microturbulence values decrease.

The values of $v_{\rm turb}$ that we find in the low chromosphere are between the upper limits ($\sim3\rm\,km\,s^{-1}$) derived from photospheric lines in plage \citep{2019arXiv190807464B}, and the larger typical values ($\sim7\rm\,km\,s^{-1}$) of nonthermal broadening implied by the optically thin \ion{O}{I} line \citep{Carlsson_2015}. Even if we correct the latter taking into account the much hotter chromospheric temperatures implied by the ALMA data, the nonthermal broadening component in the width of the \ion{O}{I} line would decrease by less than $1\rm\,km\,s^{-1}$. However, the properties of plage may significantly differ from that dataset to ours since we are only targeting the very edge of the plage region within a smaller FoV. Perhaps there are more significant spatial and/or temporal inhomogeneities of microturbulence in plage than what is commonly believed.
Unfortunately, we cannot measure this value in our IRIS data because the \ion{O}{I} line is undetected due to the short exposure time.
In any case, we note that our values would agree with the lower limits of \citep{Carlsson_2015} well within the uncertainties, and we would need greater sample sizes and  NUV spectra with higher S/N as well as higher resolution ALMA observations to further constrain nonthermal broadening.

The IRIS+ALMA inversions accentuate the differences in temperature and microturbulence between the plage region and its surroundings, which is a consequence of the spatially dependent behaviour of the response functions of the different diagnostics. ALMA has broader response functions in the hot plage regions, and narrower ones in the cool pockets.
In our models, the h and k peaks are formed higher than the bulk of the 1.25\,mm continuum everywhere outside of the plage region. However, in the hottest parts of the plage (§\,\ref{Section:shockSignatures}) the formation heights may reverse; in the weakly magnetic areas, ALMA responds to temperature perturbations around $\log\tau\sim-5$, but in the shocks the response functions have a maximum around $\log\tau\sim-5.6$ with tails extending to $\log\tau\sim-6.5$, whereas the k2v and k2r (and h2v, h2r) peaks have a contribution mostly from deeper layers ($\log\tau\sim-4.2$). There is a secondary contribution to the k2v/k2r intensity at $\log\tau\sim-6$ that evolves in time with the shock: in particular, when the 1.25\,mm continuum brightens during the shock, the k2r peak intensity increases (see Fig.\,\ref{fig:bursts}) and the relative contribution from the layers around $\log\tau\sim-6$ (that is closer to the 1.25\,mm) also increases, while the inverse happens in the k2v peak. 

Observations with ALMA Band\,3 could provide additional constraints higher up in the atmosphere. Likewise, including Band\,7 at 0.8\,mm, which is formed slightly lower than Band\,6 \citep[e.g.,][]{Wedemeyer16}, could provide constraints that could be used to resolve: (1) the problem of the UV triplet misfit that sometimes causes an exaggerated temperature increase in the lower chromosphere ($\log\tau\sim-4$) in plage, and (2): the excess microturbulent broadening in lower layers than the ones probed by Band\,6.

\subsection{Limitations of the inversions}
\label{section:caveats}
This first attempt at creating models based on fits to NUV spectra taken with IRIS and the 1.25\,mm continuum observed with ALMA was made under a few simplifying assumptions. Some of these assumptions relate to the specificities of the STiC code which assumes a 1D, plane-parallel atmosphere in hydrostatic equilibrium that is interpolated in between pre-defined nodes. Nevertheless, this approach has been shown to provide reasonable results on real data \citep[e.g.,][]{2016ApJ...830L..30D,2018ApJ...857...48G,2019A&A...627A.101V,2019ApJ...870...88E}. Other assumptions relate to the technicalities of the ALMA interferometric data that are composed of a discrete set of spatial scales unlike the UV data which show all scales down to the IRIS spatial resolution. The critical premise is that these two kinds of data can be directly compared.

The difference in spatial resolution itself of almost a factor two and the calibration uncertainties also play a role in the inversion accuracy \citep[e.g.,][]{2019arXiv190902604D}. The lower resolution of the ALMA data is not as much of a problem as it would be for IRIS because the Planck function has almost linear dependency on temperature at long wavelengths. This means that the smoothed ALMA brightness temperatures correspond to smoothed gas temperature maps, whereas smoothing the NUV would result in a more nonlinear mapping between intensity and temperature. Therefore, we chose not to sacrifice the information provided by IRIS, and we instead perform the inversions in the IRIS native resolution. 
This is nevertheless a potential source of error in the parts of the FoV where there might be sharper temperature gradients at scales shorter than the maximum ALMA resolution ($\sim$0.8").
This can only be investigated with future observations at higher resolution, or using spatially coupled inversions \citep{2019arXiv190902604D} that require a better understanding of the ALMA instrumental profile, both of which would improve the contrast at higher spatial scales. For now, a discussion on the general trends at larger spatial scales is more meaningful.

The fact that we are able to reproduce the NUV spectra and the mm wavelength point simultaneously with a relatively low error (<5\%) for most of the FoV is quite remarkable and shows that the calibration errors are probably not particularly large, or are not so large as to make this kind of inversion unfeasible. An alternative explanation is that the inferred temperatures from \ion{Mg}{II} are somewhat degenerate in the range $\log\tau\sim[-6,-5]$, meaning that they can accommodate a fairly large error in the mm intensities. 

% ------------------------------------------------ %
\subsection{The cool chromosphere}
\label{section:coolChromosphere}
The ALMA chromospheric holes stand out in the distribution of residuals as a failure of the inversions (see Fig.\,\ref{fig:fig5}). The largest of them at the top of the raster could be related to the underlying pore with a strong vertical magnetic field. Other small dark patches within the IRIS raster have no photospheric analog and are shorter-lived, but they constantly disappear and reappear in a dynamic way. There are chromospheric holes with sizes of a few arcseconds and temperatures of $\sim3000-3500$\,K that persist for several minutes outside of the FoV scanned by IRIS in the weakly magnetic areas (see Fig.\,\ref{fig:overview}). 

The inability of inversions to reproduce cool pockets in the atmosphere has been investigated by \citet{2012A&A...543A..34D} who showed that NLTE inversions of \ion{Ca}{II} 8542\,\AA~computed from 3D r-MHD simulations do not accurately predict the low-temperature regions due to 3D radiation transport effects. A similar effect is expected in the \ion{Mg}{II} lines \citep[e.g.,][]{2013ApJ...772...89L,2013ApJ...772...90L}, hence their partial visibility in intensity (Fig.\,\ref{fig:overview}).
The fact that the inversions do not perform well enough even when ALMA is added could be due to a combination of several issues such as the aforementioned non-negligible 3D radiation field in the \ion{Mg}{II} lines to which STiC is blind when given solely those lines to fit, but that turn up when ALMA tries to impose a stronger local constraint in temperature. This happens because ALMA tries to settle for a lower local electron temperature than is required by \ion{Mg}{II,} due to the scattering contribution in the source function of the latter. In addition, there could be nonequilibrium ionization effects (§\,\ref{Section:shockSignatures}).

It could also be a failure of the node parametrization scheme that is not able to capture large fluctuations in temperature over very small scale heights; in other words, the atmosphere is highly nonsmooth, as in the simulations of \citet{2012A&A...543A..34D}. We were able to improve some of the fits to the ALMA data at certain locations within the cool areas (see Fig.\,\ref{fig:Upper}) by allowing for some extra degrees of freedom, but that is not a good strategy overall. In any case, the match between the synthetic and observed mm continuum is improved when ALMA is taken into account in the inversion, which emphasizes its importance in constraining the temperature of the cool pockets that are otherwise very hard to recover. 

\citet{2019ApJ...877L..26L} recently found analogous chromospheric holes in an independent ALMA dataset in Band\,3 which suggests that these features are real. In that case, ALMA observations show that the chromosphere could harbor cool pockets where molecules such as CO could form (see Fig.\,\ref{fig:CO}). Our empirical models predict ubiquitous CO formation in the coolest regions of the photosphere and at some locations in the chromosphere except when there are hot shocks, which agrees with numerical simulations \citep[e.g.,][]{2005ESASP.596E..16W,2006ASPC..354..301W,2007A&A...462L..31W}. Therefore, cool CO clouds may be detected with the Cryogenic Near Infra-Red Spectro-Polarimeter \citep[Cryo-NIRSP,][]{2012ASPC..463..207K} to be installed on the upcoming (4m-) Daniel K. Inouye Solar Telescope, \citep[DKIST,][]{2016AN....337.1064T}.
If the chromospheric holes extend further down in height they should also be seen in ALMA Band\,7 at 0.8\,mm, which is offered from Cycle 7 onwards.

\subsection{The hot chromosphere}
\label{Section:shockSignatures}

Contrary to the chromospheric holes mentioned in the previous section, we managed to obtain very good quality fits in the plage. This is probably because at these hot locations the temperature stratification is relatively smoother in height, and the h and k lines are more coupled to the local conditions as shown by the good correspondence between bright features in IRIS and ALMA (Fig.\,\ref{fig:bursts}).
We found that the inversions with ALMA result in typically higher temperatures and lower microturbulence in the chromosphere throughout the duration of the transient brightenings in plage, unlike the inversions of IRIS data alone which infer nearly constant temperatures but increasing turbulent broadening within the 9000\,K areas of the ALMA data (see Fig.\,\ref{fig:bursts2}). The photospheric properties, which are constrained by IRIS, do not significantly change during the time span.

The fact that the brightness is increased in the magnetic regions (except above the dark patch), suggests that the magnetic field may play a role in explaining the bright features either by means of reconnection or by increasing the dissipation rate of waves.
The latter scenario gains support from the swings between blue and red peak emission in the h and k lines (Fig.\,\ref{fig:Spectra evolution}) which are well-known signatures of (slow) shock waves propagating upwards in magnetized regions \citep[e.g.,][]{2006ApJ...647L..73H,2006ApJ...648L.151J,2007ApJ...655..624D,2008ApJ...679L.167L} that heat up the plasma and shift the formation height of the 1.25\,mm continuum to higher layers. In simulations, shocks also manifest as enhancements in the mm continua of the order of 1000 K \citep{2007A&A...471..977W}. In that case the observed brightness temperatures of nearly 11\,000\,K may be evidence of nonequilibrium hydrogen ionization according to the simulations of \citet{2007A&A...473..625L} and \citet{2016ApJ...817..125G}. Following the same interpretation, the chromospheric holes (except the one above the pore) may be the mm counterparts of the cool intershock regions that reach temperatures as low as 2500\,K in simulations. 
The shock wave pattern in the \ion{Mg}{II} lines has periods of between 3.5 and 4\,min (<5\,mHz), but the changes in the mm continuum do not seem periodic (Fig.\,\ref{fig:shocks}). 
The visibility of the shock waves in the ALMA and IRIS data will be studied in more detail in a forthcoming publication (Chintzoglou et al, in prep.).

As a corollary, our statistical equilibrium calculations probably overestimate the electron densities during the shocks and give an exaggerated temperature increase with height since in non-equilibrium conditions the ionization degree of hydrogen (the major donator of electrons in the chromosphere) stays fairly constant due to the long recombination timescales. 
Detailed comparisons between 3D r-MHD simulations of plage regions taking into account time-dependent hydrogen ionization are needed to properly assess these effects on the modeled intensities, but these are beyond the scope of this paper.
% ------------------------------------------------------- %
\section{Conclusions}
\label{section:conclusion}

IRIS and ALMA data have been used for the first time to constrain the thermodynamical properties of the solar atmosphere at the vicinity of a plage region near disk center using a NLTE inversion code. 
Our findings are broadly consistent with the predictions based on the publicly available snapshots of state-of-the-art 3D MHD simulations of the solar chromosphere \citep{2016A&A...585A...4C} made by \citet{PaperI}, and they reinforce the synergy between IRIS and ALMA. 

The ALMA maps display a broad dynamical range in brightness temperature from $\sim3000$\,K in the cool patches, which we refer to as chromospheric holes, to $\sim11\,000$\,K compact brightenings that evolve on scales of only a few minutes. We analyzed examples of both of these features and conclude that they are probably purely chromospheric phenonomena because they lack obvious counterparts in the photospheric diagnostics, which is supported by our inversion results. While the chromospheric holes could be caused by a sharp drop in temperature at low continuum optical depths where cool CO clouds could form, the brightenings seem to be associated with a hot propagating shock wave and possibly nonequilibrium ionization effects \citep[e.g.,][]{2007A&A...473..625L}, both of which require nonsmooth temperature stratifications that are difficult to infer based on inversions of NLTE lines alone. 

Overall, including the 1.25\,mm continuum in inversions of IRIS NUV data sets stronger constraints on temperatures and microturbulences around $\log\tau\sim-5$ and lower optical depths where the response functions peak. The ALMA data show that the temperatures in the chromosphere of plage are on average $\sim8500$\,K, that is $\sim2000$\,K hotter than what the \ion{Mg}{II} lines suggest \citep{Carlsson_2015}, and this cannot be exclusively attributed to the uncertainty in the ALMA flux calibration. Interestingly, the IRIS+ALMA inversions tend to infer lower microturbulence: $\sim$2$\rm\,km\,s^{-1}$ is the median in the whole FoV compared to $\sim$4$\rm\,km\,s^{-1}$ inferred from IRIS alone. In the shocks in the plage region the 1.25\,mm is formed at lower optical depths than in the weakly magnetized areas (around $\log\tau\sim-5.6$), and the depth- or spatially averaged microturbulence is $\sim5(\pm1)\rm\,km\,s^{-1}$, which is similar to what is predicted by the IRIS inversions but with a skew to higher values in the latter. Our values are within the lower limits of nonthermal broadening in the \ion{O}{I} lines \citep{Carlsson_2015}, but differences in formation heights as well as spatial and temporal inhomogeneities of microturbulence have to be taken into account. ALMA observations of plage regions at higher spatial resolution and IRIS FUV spectra with  higher S/N and larger rasters are needed to further investigate these findings.

Whether other lines (even photospheric diagnostics) in the IRIS spectral windows can be included in the inversions remains to be investigated. The results described in the above highlight the need for more work on inversions that combines data from more lines such as the \ion{C}{II} lines in the IRIS FUV passband, as well as other NLTE lines such as \ion{Ca}{II} H, K, and 8542\,\AA~obtained from ground-based telescopes like SST and DKIST. 

The \ion{Mg}{II} lines alone may not be sufficient to infer accurate chromospheric temperatures and microturbulence due to well-known NLTE effects, but they greatly help us to interpret ALMA observations because they provide complementary information on temperature, LOS velocities, and microturbulence \citep[e.g.,][]{2016ApJ...830L..30D}. Furthermore, fast inversions of the h, k, and UV triplet lines are now easily available through the \textit{IRIS}${}^2$ database \citep{2019ApJ...875L..18S}.
This paper shows that the combination of IRIS and ALMA is a powerful tool for diagnosing a wider range of physical conditions in the atmosphere, and it underscores the need for more coordinated observations and 3D r-MHD simulations of plage and other solar features taking into account nonequilibrium hydrogen ionization to better understand the formation of the IRIS and ALMA diagnostics in chromospheric conditions.

\begin{acknowledgements}
This paper makes use of the following ALMA data: ADS/JAO.ALMA\#2016.1.00050.S. ALMA is a partnership of ESO (representing its member states), NSF (USA) and NINS (Japan), together with NRC (Canada), MOST and ASIAA (Taiwan), and KASI (Republic of Korea), in cooperation with the Republic of Chile. The Joint ALMA Observatory is operated by ESO, AUI/NRAO and NAOJ.

IRIS is a NASA small explorer mission developed and operated by LMSAL with mission operations executed at NASA Ames Research center and major contributions to downlink communications funded by ESA and the Norwegian Space Centre.

Hinode is a Japanese mission developed and launched by ISAS/JAXA, with NAOJ as domestic partner and NASA and STFC (UK) as international partners. It is operated by these agencies in co-operation with ESA and NSC (Norway).

The Institute for Solar Physics is supported by a grant for research infrastructures of national importance from the Swedish Research Council (registration number 2017-00625).

The computations were performed on resources provided by the Swedish National Infrastructure for Computing (SNIC) at the High Performance Computing Center North (HPC2N) at Umeå University, and the National Supercomputer Centre (NSC) at Linköping University.

JdlCR is supported by grants from the Swedish Research Council (2015-03994), the Swedish National Space Board (128/15) and the Swedish Civil Contingencies Agency (MSB). This project has received funding from the European Research Council (ERC) under the European Union's Horizon 2020 research and innovation programme (SUNMAG, grant agreement 759548). JL was supported by a grant (2016.0019) from the Knut and Alice Wallenberg foundation. BDP and GC were supported by NASA contract NNG09FA40C (IRIS).

This work is supported by the SolarALMA project, which has received funding from the European Research Council (ERC) under the European Union’s Horizon 2020 research and innovation programme (grant agreement No. 682462), and by the Research Council of Norway through its Centres of Excellence scheme, project number 262622.

This work benefited from discussions within the activities of team 399 "Studying magnetic-field-regulated heating in the solar chromosphere" at the International Space Science Institute (ISSI) in Switzerland. 

This research has made use of SunPy, an open-source and free community-developed solar data analysis Python package \citep{2015CS&D....8a4009S}.

\end{acknowledgements}

\bibliographystyle{aa} % style aa.bst
%\bibliography{bib.bib} % your references Yourfile.bib

\end{document}